\documentclass[
aip,
amsmath,amssymb,
reprint,%
]{revtex4-1}
\usepackage[colorlinks=true,
linkcolor=red,
urlcolor=black,
citecolor=blue]{hyperref}
\usepackage{graphicx}
\usepackage{epstopdf}
\usepackage{dcolumn}
\usepackage{bm}
\usepackage{soul}

\usepackage{color}
\definecolor{darkgreen}{rgb}{0.1,.6,.1}
\definecolor{greenblue}{rgb}{0.0,.1,.4}
\usepackage[normalem]{ulem}

\begin{document}


\title {Transition to hyperchaos and rare large-intensity pulses in Zeeman laser}
\author{S. Leo Kingston}
\thanks{kingston.cnld@gmail.com}
\affiliation{Division of Dynamics, Lodz University of Technology,   90-924 Lodz, Poland}	
\author{Marek Balcerzak}
\affiliation{Division of Dynamics, Lodz University of Technology,   90-924 Lodz, Poland}
\author{Syamal K. Dana}
\affiliation{Division of Dynamics, Lodz University of Technology,   90-924 Lodz, Poland}
\affiliation{Department of Mathematics, National Institute of Technology, Durgapur 713209, India}
\author{Tomasz Kapitaniak}
\affiliation{Division of Dynamics, Lodz University of Technology,  90-924 Lodz, Poland}

\date{\today}



\date{\today}

\begin{abstract}
A discontinuous transition to hyperchaos is observed at discrete critical parameters in the Zeeman laser model for  three well known nonlinear sources of instabilities, namely, quasiperiodic breakdown to chaos  followed by interior crisis,  quasiperiodic intermittency, and Pomeau-Manneville intermittency. Hyperchaos appears with a sudden expansion of the attractor of the system at a critical parameter for each case and it coincides with triggering of occasional and recurrent large-intensity pulses. The transition to hyperchaos from a periodic orbit via Pomeau-Manneville intermittency shows hysteresis at the critical point, while no hysteresis is recorded during the other two processes.  
The recurrent large-intensity pulses show characteristic features of  extremes by their height  larger than a threshold and probability of rare occurrence. The phenomenon is robust to weak noise although the critical parameter of transition to hyperchaos shifts with noise strength. This phenomenon appears as common in many low dimensional systems  as reported earlier \cite{chowdhury2022extreme}, there the emergent large-intensity events or extreme events dynamics have been  recognized simply as chaotic in nature although the temporal dynamics shows occasional large deviations from the original chaotic state in many examples. We need a new metric, in the future,  that would be able to classify such significantly different dynamics and distinguish from chaos.

\end{abstract}

\maketitle
\begin{quotation}
A Zeeman laser model has been investigated here, which shows a variety of instabilities that are usually found in other systems. However, we report the origin of extremely large-intensity pulses via quasiperiodic breakdown to chaos, quasiperiodic intermittency, Pomeau-Manneville intermittency, and their intrinsic relation with the hyperchaotic dynamics. Intriguingly, the transition to large-intensity pulses and the hyperchaotic dynamics appear concurrently, which is confirmed by the existence of two positive Lyapunov exponents in the system. Distinct transitions to large-intensity pulses in the Zeeman laser model are robust with weak noise.    
\end{quotation}
\section{Introduction}
Haken \cite{haken1975analogy} first reported chaos in Maxwell-Bloch equations similar to what was found in the atmospheric circulation model proposed by Lorenz \cite{lorenz1963deterministic}. This  encourages the optical community to explore various intriguing nonlinear phenomena including chaos in  deterministic and stochastic laser systems \cite{ohtsubo2012semiconductor,lundqvist2013order,arecchi1988instabilities,tredicce1985instabilities,narducci1988laser,mandel2005theoretical,narducci1988laser,mandel2005theoretical}. Chaos is characterized by directional exponential stretching of the trajectory of a nonlinear dynamical system, which is detected by one positive Lyapunov exponent \cite{alligood1996chaos} of the system. Existence of two positive Lyapunov exponents was first identified in a four-dimensional system by R\"ossler \cite{rossler1979equation} and defined as hyperchaos.  Hyperchaos was later observed in a variety of systems, electronic circuits\cite{kapitaniak1993transition, kapitaniak1994hyperchaotic,kengne2015coexistence}, coupled oscillators \cite{perlikowski2010routes,stankevich2018chaos, stankevich2021chaos}, finance systems \cite{szuminski2018integrability}, radiophysical generator \cite{stankevich2019chaos},  interacting gas bubbles in a liquid \cite{garashchuk2019hyperchaos},  and two element nonlinear chimney model\cite{kashyap2020hyperchaos}. It has also been reported in lasers \cite{ahlers1998hyperchaotic,bonatto2018hyperchaotic, colet1994controlling, roy2020synchronization}.  Control of hyperchaos was implemented in experiments using occasional proportional feedback in multimode Nd:YAG laser \cite{colet1994controlling}. Synchronization of hyperchaotic systems was investigated in semiconductor lasers \cite{ahlers1998hyperchaotic}.    
\par In recent time, another class of complex dynamics such as the rare extremely large-intensity events (LIE) has been investigated in  laser systems, $\mathrm{CO_2}$ lasers \cite{bonatto2017extreme}, semiconductor laser\cite{reinoso2013extreme,  mercier2015numerical} and microcavity laser \cite{coulibaly2017extreme}. In particular, studies on rouge waves in lasers  \cite{solli2007optical, lecaplain2014rogue, bonatto2011deterministic, pisarchik2011rogue} made important inroads in extreme events' research \cite{mishra2020routes,chowdhury2022extreme}. Extreme events have characteristic features more complex than the classical chaotic motion in their dynamical as well as statistical properties. Localized  dynamical instabilities \cite{chowdhury2022extreme,mishra2020routes},  namely,  interior-crisis-induced intermittency \cite{  kingston2017extreme, ray2019intermittent,kingston2020extreme,ouannas2021chaos}, Pomeau-Manneville (PM) intermittency \cite{pomeau1980intermittent, kingston2017extreme, kingston2021intermittent, mishra2018dragon}, torus breakdown of quasiperiodic motion \cite{mishra2020routes, kingston2021instabilities}, and quasiperiodic intermittency \cite{kingston2021instabilities}, are involved  in the triggering of extreme events in a wide variety of nonlinear systems. Noise-induced attractor hopping in multistable system \cite{pisarchik2011rogue}, and instability of in-phase \cite{cavalcante2013predictability} and antiphase synchronization \cite{saha2017extreme, mishra2018dragon} in coupled systems can also originate extreme events. In the process, rare and recurrent large amplitude events  show up in  dynamical systems, in general, when the trajectory of the system leaves its bounded phase space, on rare occasions, to travel far away distance, but returns to the original location after a while. The short-lived occasional journey to far away location have been triggered by the local instabilities in phase space of the system and originate large amplitude events that are called as extremes when they are larger than a threshold height and rare in occurrence. 
\par Extreme events \cite{bonatto2017extreme} and hyperchaos \cite{bonatto2018hyperchaotic} have been studied in lasers, separately, in search of  complexity of dynamics, however, any link between the two phenomena has been  unnoticed. Indeed, they represent two sides of the same coin that we reveal here in our numerical study of the seven-dimensional Zeeman laser model \cite{redondo1997intermittent,redondo1996off}. In particular, we consider this laser model, for our study, because it shows three different sources of instabilities that lead to recurrent and rare large-intensity pulses as reported earlier \cite{kingston2021instabilities}. There it was shown that rare large-intensity events appear at two different critical parameters in the Zeeman laser model mainly via two nonlinear processes,  quasiperiodic breakdown to chaos followed by interior crisis, and quasiperiodic intermittency. In addition, we report here  triggering of LIE via PM intermittency in the laser model as found in our recent search for other sources of instabilities that may exist in the broad parameter space of the system. The typical PM intermittency  \cite{manneville1979intermittency} is found at another critical parameter when the turbulent phase (chaotic bursting) occasionally intercepts the laminar phase (almost periodic state). In the turbulent phase, the height of the chaotic bursts may be larger than a threshold height. All the three sources of nonlinear instabilities trigger a discontinuous large expansion of the attractor as manifested in the bifurcation diagrams of the system at discrete critical parameters along with the emergence of hyperchaos.  In reality, the expansion of the attractor is not permanent, but intermittent when the trajectory of the system occasionally visits far away location of state space of the system originating short-lived LIE, which are recurrent and rare in occurrence. 
\par Hyperchaos was observed earlier \cite{bonatto2018hyperchaotic} in a free-running laser diode  with a variety of possible statistical distribution of light-polarized dynamics. Very recently, hyperchaos is reported in semiconductor superlattice \cite{mompo2021designing} that is smaller in size and equally fast compared to optical devices. It was clearly shown there that a sudden transition to large amplitude oscillation occurs at a critical parameter via PM intermittency. The temporal dynamics shows intermittent large amplitude events, however, the specific characteristic features of the temporal dynamics was not discussed there. We present here  two more sources of instabilities, besides the PM intermittency, that originate hyperchaos with concurrent appearance of LIE in the Zeeman laser. Interestingly, we observe a common scenario for all the three sources of instabilities (quasiperiodic breakdown to chaos followed by crisis, PM intermittency and quasiperiodic intermittency) that lead to a discontinuous large expansion of the attractor of the system at discretely different critical parameters with the origin of hyperchaos and rare LIE. In fact, such processes of simultaneous origin of hyperchaos and LIE are common as reported earlier \cite{kingston2022transition} in other paradigmatic models.  
 \par In many low dimensional models, including lasers, a sudden large expansion of the attractor of the system was shown with the origin of extreme events. However, such extreme events were recognized as simple chaos although they were more complex in nature.  It has not been explored so far, to the best of our knowledge, how complex really they are in comparison to chaos? Possibly, we need a different metric to classify them from simple chaos.  For higher dimensional systems, it becomes easier to classify extreme events as hyperchaotic. To prove the statement, we consider the 7-dimensional Zeeman that shows three varieties of dynamical mechanisms leading to extreme events  denoted here as LIE.  The sources of instabilities are first located in  parameter space of the laser model and then the origin of hyperchaos is identified in response to a change in a system parameter. The Lyapunov exponents of the system  are estimated using a perturbation method \cite{kingston2021instabilities, balcerzak2018fastest} to identify the emergence of hyperchaos. We  draw bifurcation diagrams of the system to locate the critical parameter point of large expansion of the attractor and the appearance of LIE and then confirm the statistical properties of extreme heights and rare occurrence.  We have also checked whether the discontinuous transition to hyperchaos against the parameter is hysteresis free or not by doing forward and backward integration. This phenomenon of the origin of hyperchaos with a discontinuous expansion of the attractor at a critical parameter and the emergence of rare LIE is robust to weak noise, although the critical parameter for transition to hyperchaos shifts against the noise strength.
 
  \par We like to mention that this article is presented here in recognition of the fundamental contributions made by Prof. J\"urgen Kurths in the field of nonlinear dynamics, synchronization of chaos and dynamical networks including climate networks, geophysical sciences, and as a tribute to commemorate his 70th birthday which falls in March 2023.
\section {Zeeman laser model}
The emergence of rare and recurrent LIE have already been reported by Kingston $et~al.$ \cite{kingston2021instabilities} in the Zeeman laser model \cite{redondo1997intermittent,redondo1996off},	
\begin{eqnarray}
\dot{E}_x &=&\sigma(P_x - E_x), \\ \nonumber
\dot{E}_y &=&\sigma(P_y - \alpha E_y), \\ \nonumber
\dot{P}_x &=&-P_x +E_x D_x+E_y Q, \\  \nonumber
\dot{P}_y &=&-P_y +E_y D_y+E_x Q, \\  \nonumber
\dot{D}_x &=&(r-D_x)-2(2E_x P_x+E_y P_y), \\  \nonumber
\dot{D}_y &=&(r-D_y)-2(2E_y P_y+E_x P_x), \\  \nonumber
\dot{Q} &=&-Q-(E_x P_y+E_y P_x),   \nonumber
\label{Zee:eqn}
\end{eqnarray}	
\begin{figure}
\includegraphics[width=0.49\columnwidth]{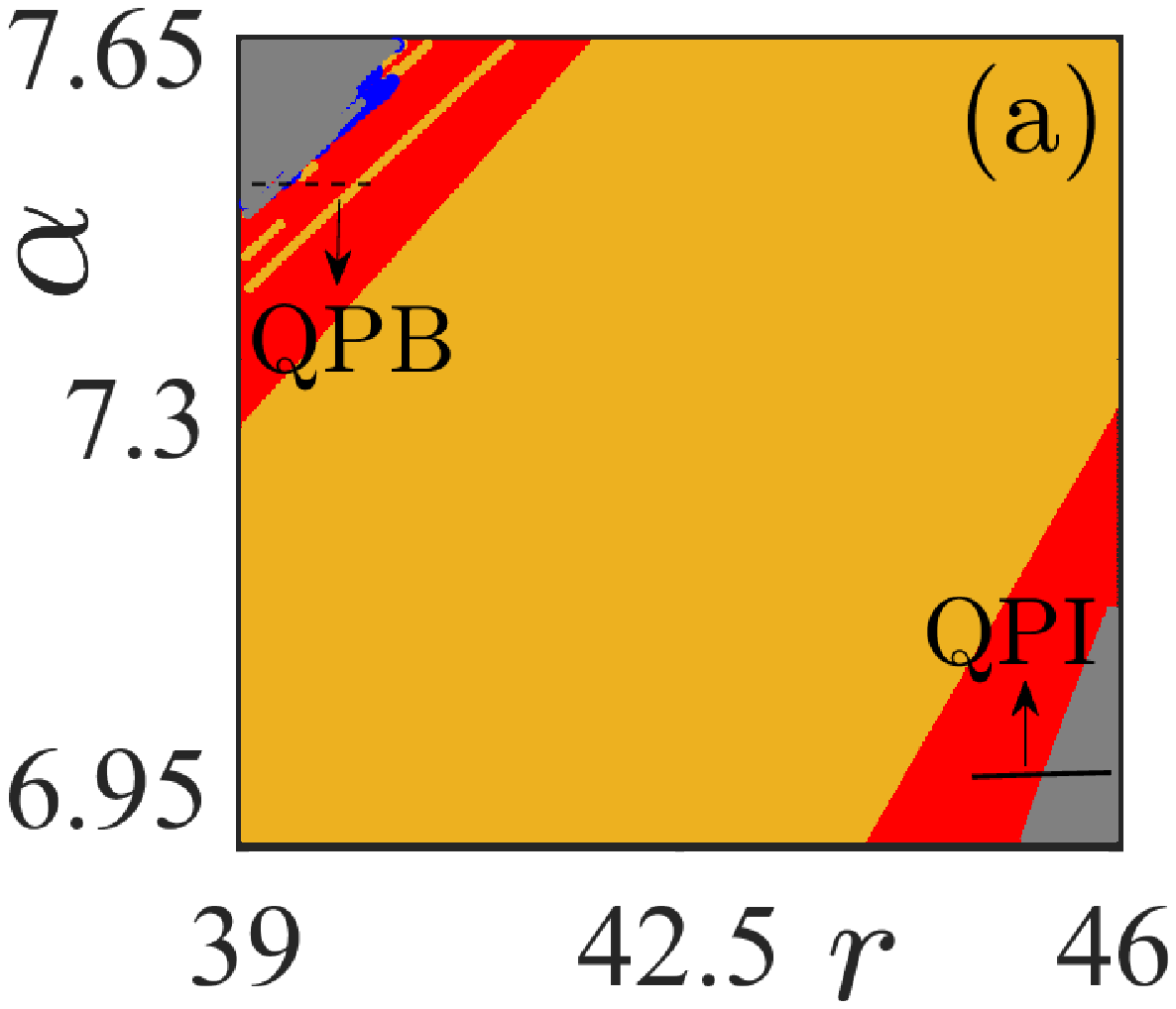}~
\includegraphics[width=0.5\columnwidth]{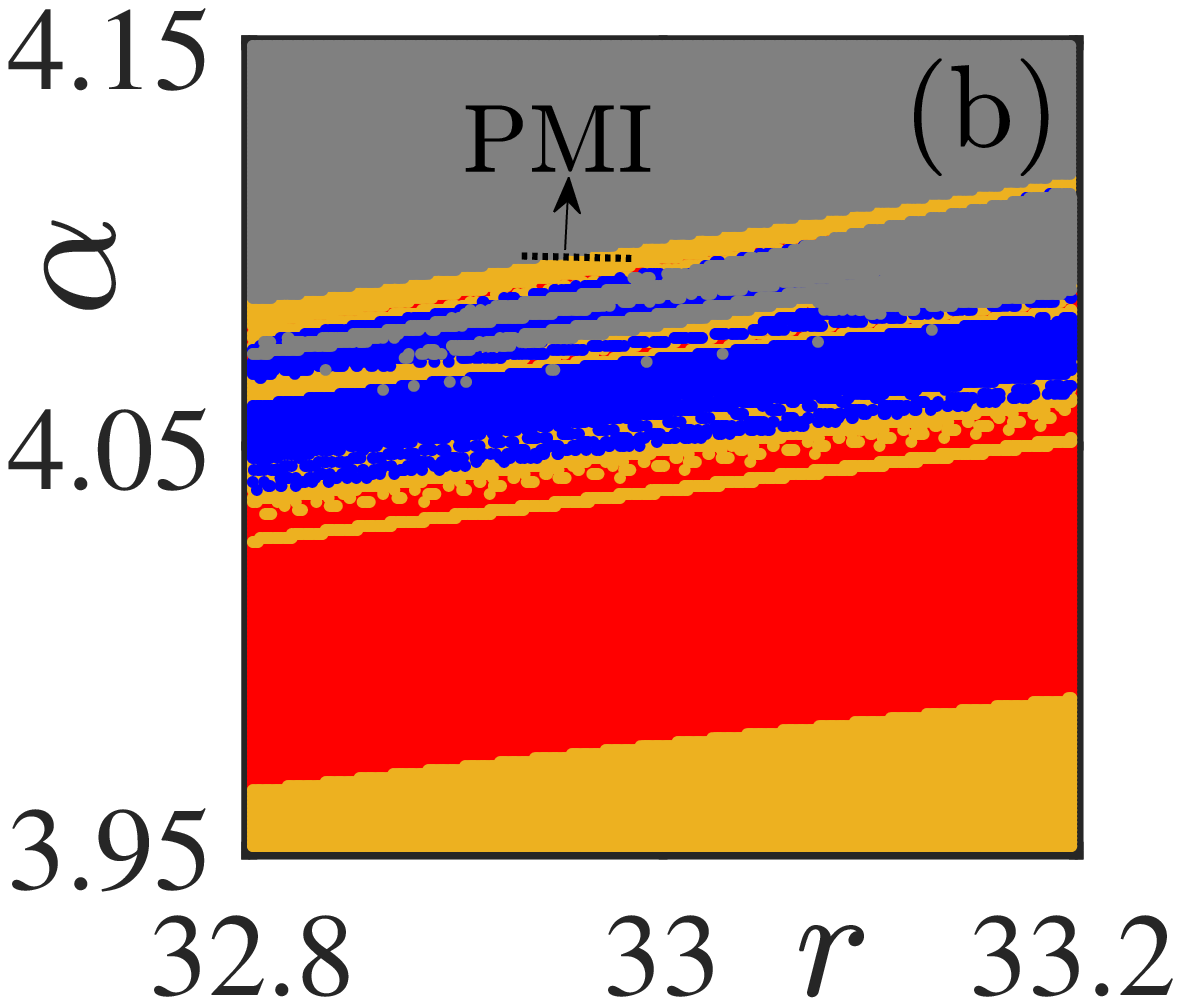}	
	\caption{Phase diagrams in a $r$ - $\alpha$ plane of the Zeeman laser model.  Dynamical states, periodic (yellow), quasiperiodic (red), chaotic (blue), and hyperchaotic (gray) are shown. (a) Transition to hyperchaos via QPB to chaos, and QPI are identified in two regions as marked by horizontal dashed black line and solid black line, respectively. (b) Transition to hyperchaos via PMI is identified in a region marked by a dotted line.  Parameter $\sigma=6.0$.}
	\label{fig1}
\end{figure}
\begin{figure}
	\includegraphics[width=0.85\columnwidth]{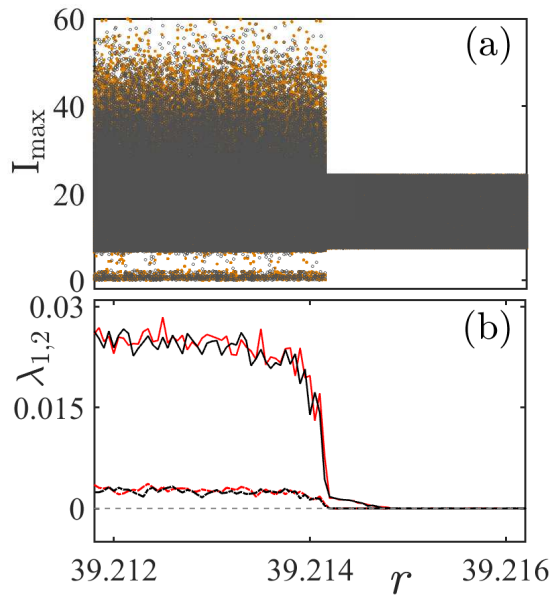}\\	
	\caption{Quasiperiodic breakdown to chaos followed by hyperchaos. Bifurcation diagrams against $r \in (39.2115, 39.2165)$, (a) local maxima  $I_{max}$ of laser intensity is plotted against $r$ in two colors (yellow dots and gray circles)  indicating forward and backward intregations and (b) two largest Lyapunov exponents $\lambda_{1,2}$ against $r$ are plotted in  two colored lines (red and black lines)  indicating forward and backward integrations. A  sudden  large expansion of $I_{max}$ is seen (a) at a critical point $r$ = $39.2141$  when the second largest Lyapunov exponent $\lambda_2$ transits to a positive value (b) indicating origin of hyperchaos since $\lambda_1$ already becomes positive at $r=39.2144$.  The transition to hyperchaos  triggers LIE  at critical  $r =39.2141$ as illustrated in the time evolution of $I$ in Fig.~\ref{fig3}(d). Parameters $\sigma=6.0$, $\alpha=7.51$.}
	\label{fig2}
\end{figure}
\begin{figure}
\includegraphics[width=0.52\columnwidth]{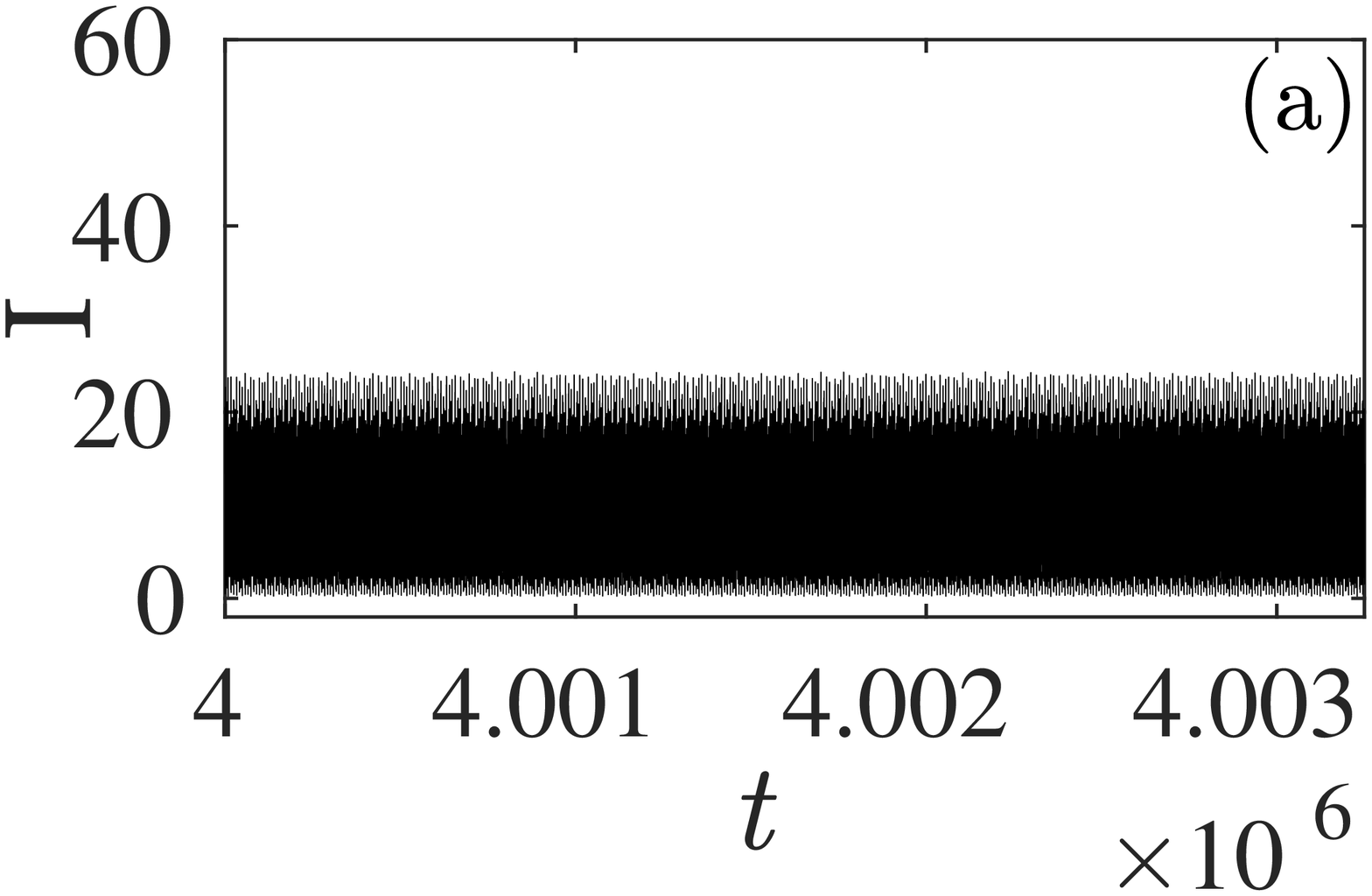}~\includegraphics[width=0.52\columnwidth]{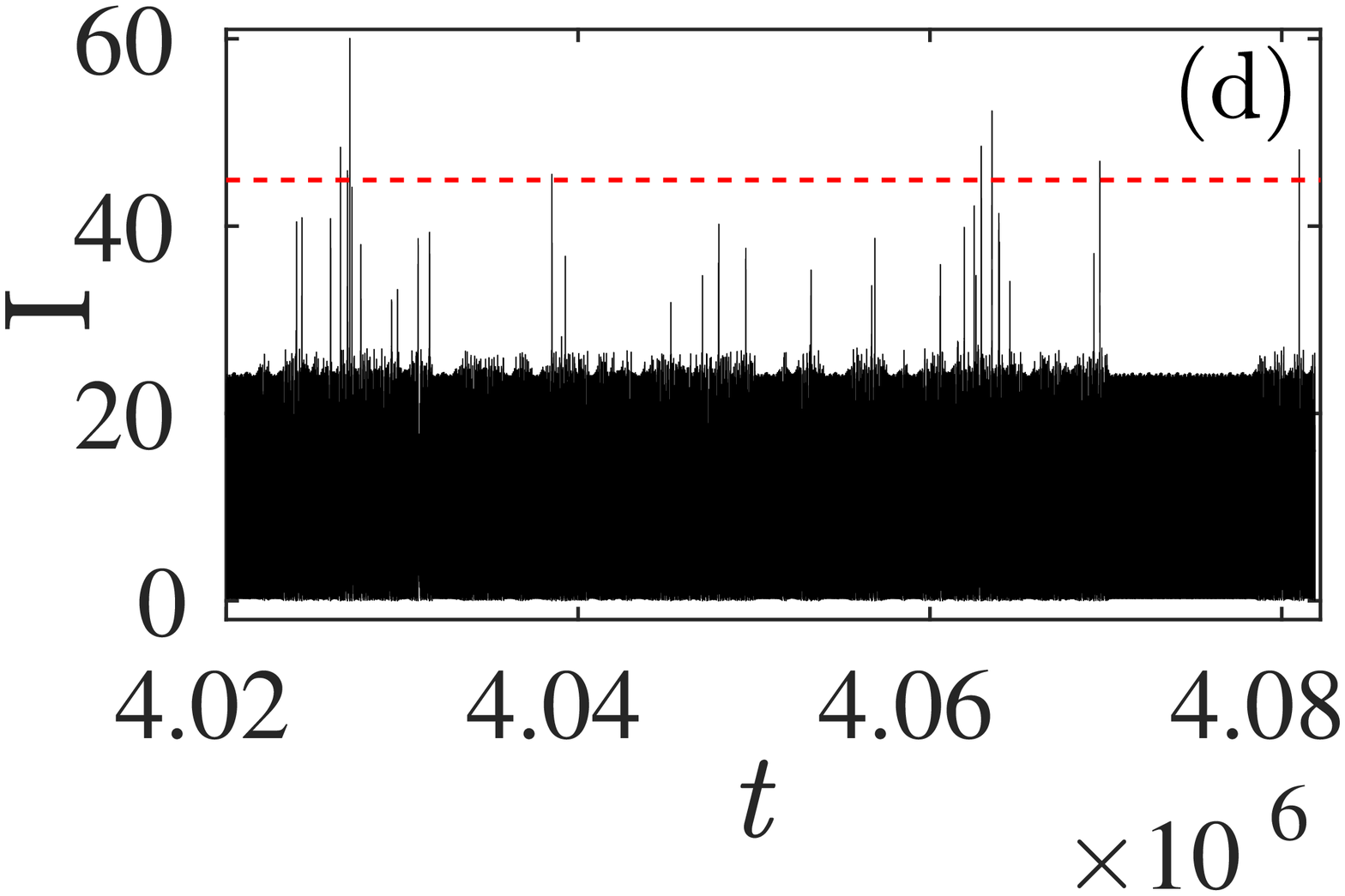}\\	

\includegraphics[width=0.52\columnwidth]{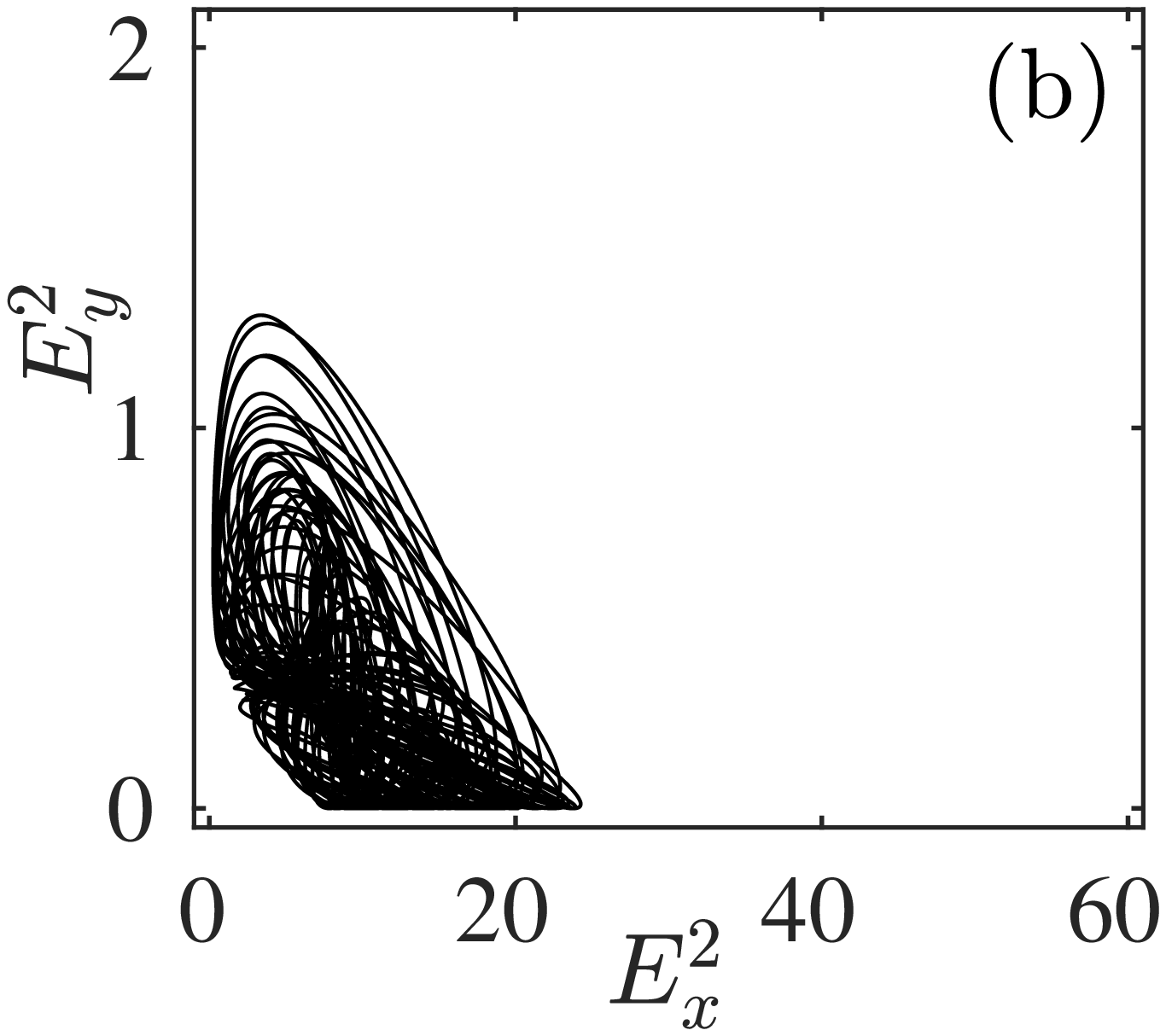}~\includegraphics[width=0.52\columnwidth]{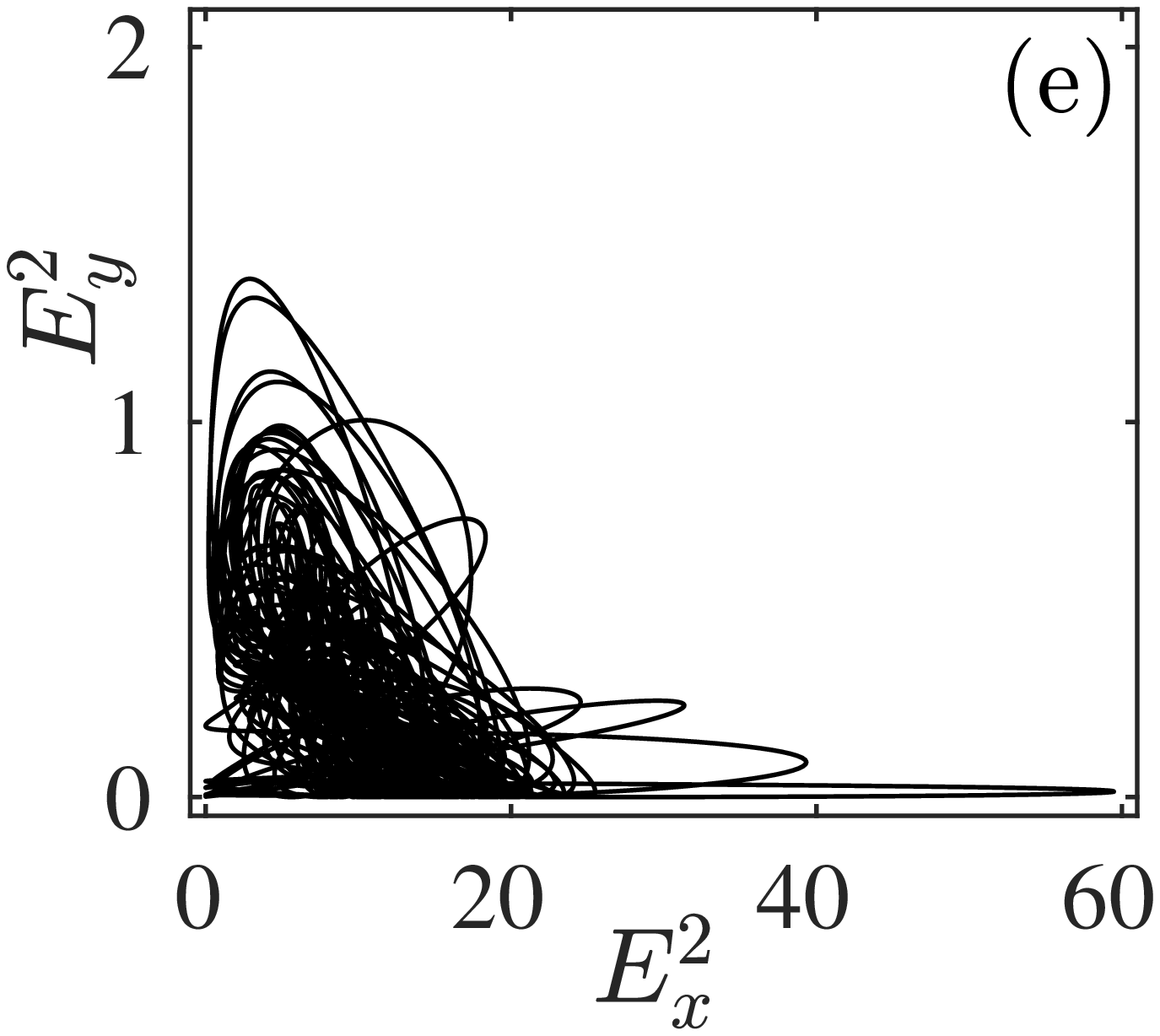}

\includegraphics[width=0.52\columnwidth]{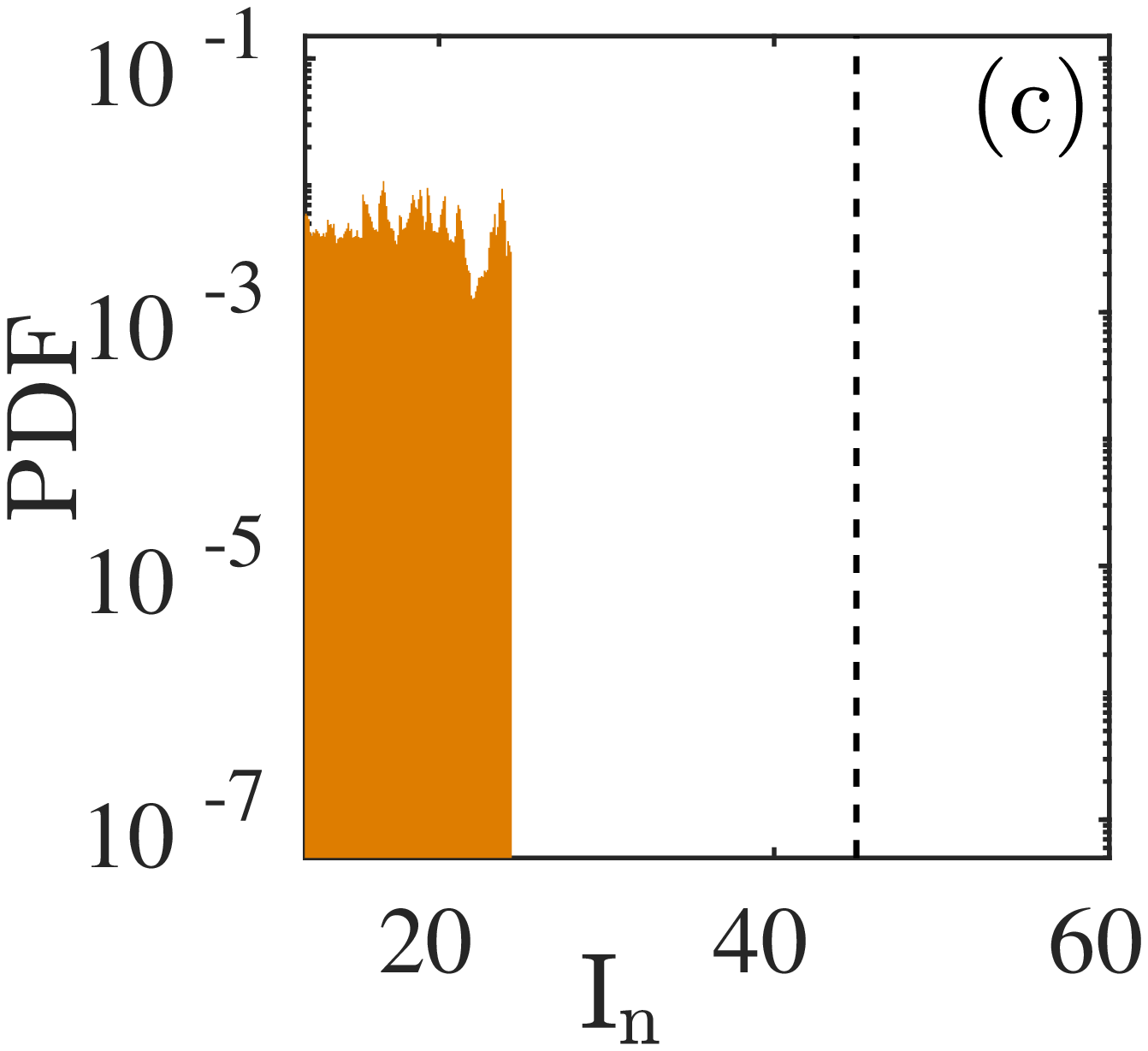}~\includegraphics[width=0.52\columnwidth]{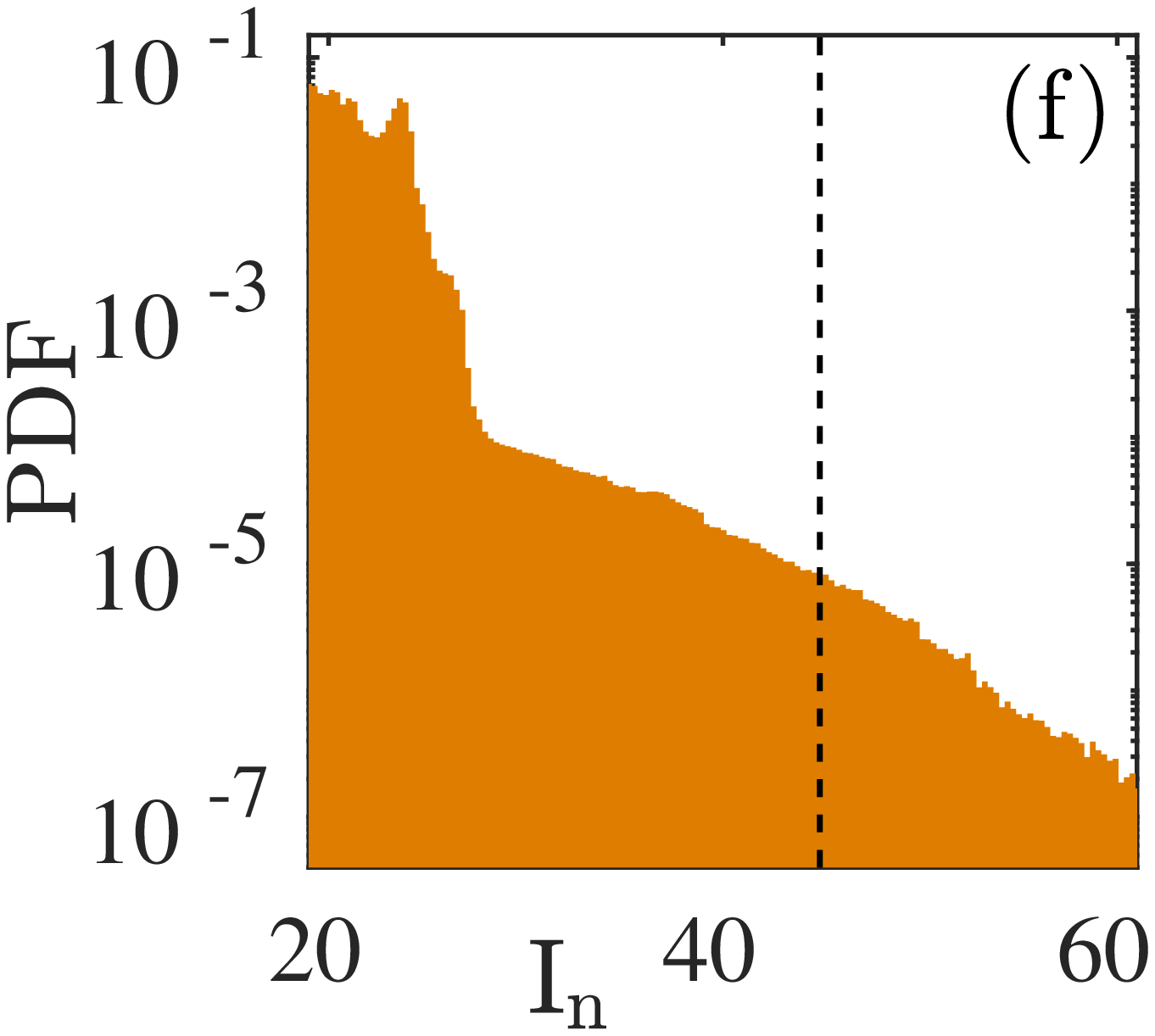}

\includegraphics[width=0.525\columnwidth]{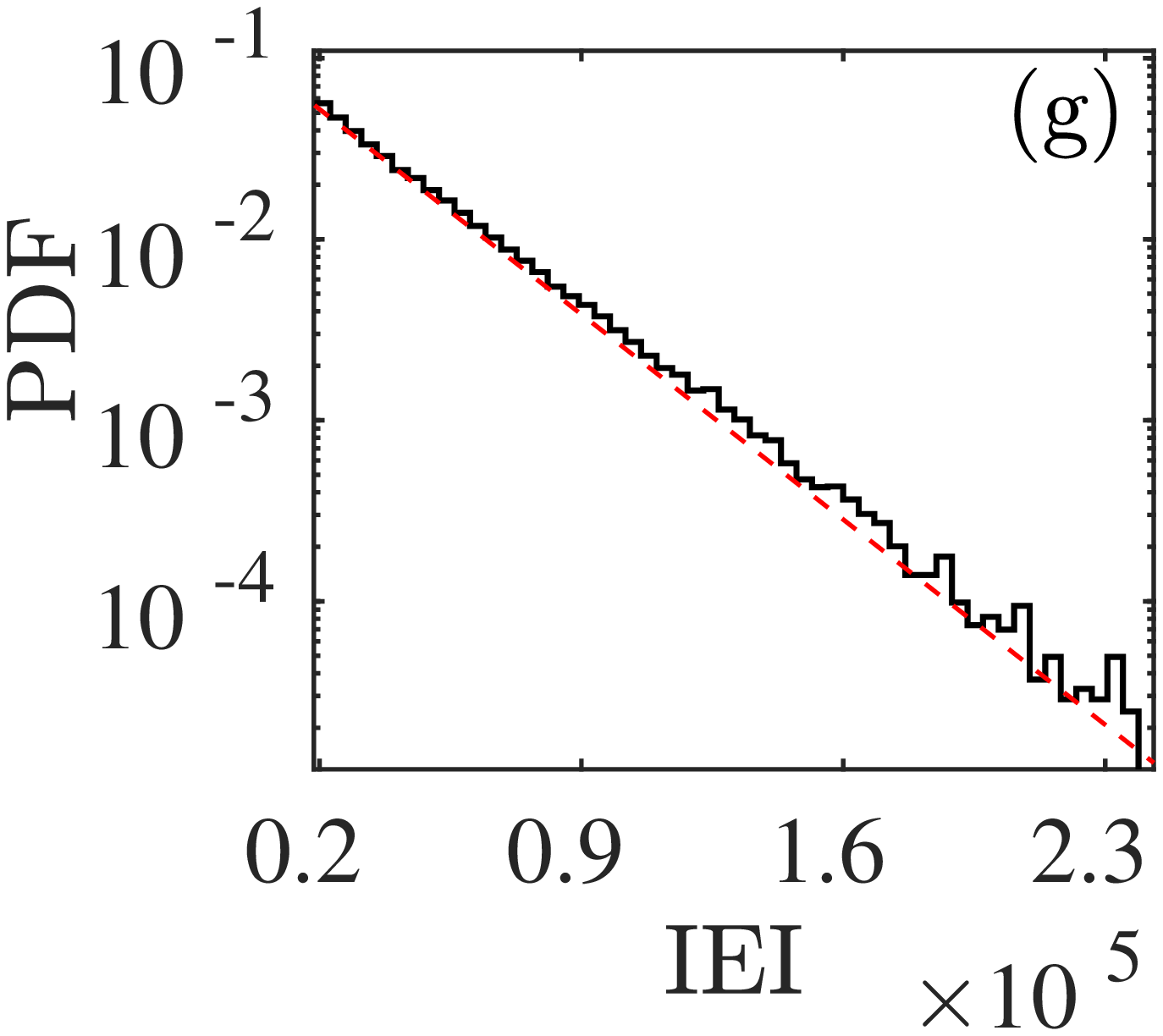}
	\caption{Hyperchaos and large-intensity pulses via quasiperiodic breakdown. (a) Time evolution of $I$ in chaotic state for $r \approx$  39.2142, and  its phase portrait $E_x^2$ against $E_y^2$ in (b), and (c) probability distribution that shows bounded below the threshold line (vertical dashed line). (d) LIE with occasional crossing of the significant height $h_s$  marked by a horizontal dashed red line for $r=39.2141$. (e) Phase portrait $E_x^2$ against $E_y^2$ shows occasional large event away from the dense region in phase space, and the PDF of LIE in (f) shows an extended tail. The $h_s$ line is marked by vertical dashed line in (f). (g) Distribution of IEI fits (red dashed line) well with an exponential function. }
	\label{fig3}
\end{figure}
where $E_x$ and $E_y$  are the state variables representing the linear polarization components of the electric field, $I=E^2_{x}+E^2_{y}$ is the laser intensity, and  ($P_x, P_y$) and ($D_x, D_y$) are proportional to the polarization and atomic inversion, respectively, and related to a transition $| J=1, J_i = 0\rangle \leftrightarrow |J=0\rangle$, and $Q$ is proportional to the coherence between the upper sub levels $| J=1, J_x = 0\rangle$ and $| J=1, J_y = 0\rangle$.  The parameter $r$ denotes the incoherent pumping rate, and $\sigma$ and $\alpha \sigma$ represent the cavity losses along the $x$ and $y$ directions, where $\alpha$ is the cavity anisotropy parameter.  
\par We search the broad parameter region in a $(r-\alpha)$ plane of the model to locate the sources of instabilities, as usual, that trigger the LIE  and then present
two phase diagrams in Fig.~\ref{fig1}. Four different dynamical regimes have been identified  such as periodic (yellow), quasiperiodic (red), chaos (blue), and hyperchaos (gray) that are delineated  by using the largest Lyapunov exponents of the model. 
The regions of transitions are marked to locate  quasiperiodic breakdown (QPB, horizontal dashed line) to chaos, quasiperiodic intermittency (QPI, solid horizontal line) in Fig.~\ref{fig1}(a), and  PM intermittency (PMI, short horizontal dotted line) in Fig.~\ref{fig1}(b). 
\section{Hyperchaos: Quasiperiodic breakdown}

For a demonstration of the origin of hyperchaos via  breakdown of quasiperiodicity followed by an interior crisis, the pumping parameter $r$ is varied along the dashed black line in Fig.~\ref{fig1}(a) and then we plot a bifurcation diagram of local maxima $I_{max}$ against $r$ in Fig.~\ref{fig2}(a) in a range of $r \in (39.2115, 39.2165)$.  A sudden large expansion of $I_{max}$ is seen at a critical $r=39.2141$,  but the chaotic regime is not clear from the bifurcation diagram. However, the Zeeman laser is known to exhibit breakdown of quasiperiodicity via successive torus-doubling to chaos as reported recently \cite{kingston2021instabilities}. To check the transition to chaos, two largest Lyapunov exponents ($\lambda_{1,2}$) of the system are drawn in  Fig.~\ref{fig2}(b) that clearly illustrates the origin of chaos at a critical point. The $\lambda_{1,2}$ continue to be zero for a range of decreasing $r$ value confirming quasiperiodic motion and then the first largest Lyapunov exponent $\lambda_1$ becomes positive at a critical $r=39.2144$ indicating origin of chaos when the second Lyapunov exponent $\lambda_2$ remains zero. The existence of chaos continues for a small range of $r$ until at $r=39.2141$, and the second largest Lyapunov exponent $\lambda_2$ becomes positive that confirms the origin of hyperchaos. The sudden large expansion of $I_{max}$  coincides with the origin of hyperchaos as manifested in Fig.~\ref{fig2}(a). Both forward and backward integrations of the laser model are done and plotted in Figs.~\ref{fig2}(a)-(b) (marked by two different colors) and we find no shift in the transition point in both the large expansion of $I_{max}$ and the arrival of hyperchaos (transition of $\lambda_2$ to a positive value) that negate the presence of any hysteresis during this transition. 
\begin{figure}   
	\includegraphics[width=0.85\columnwidth]{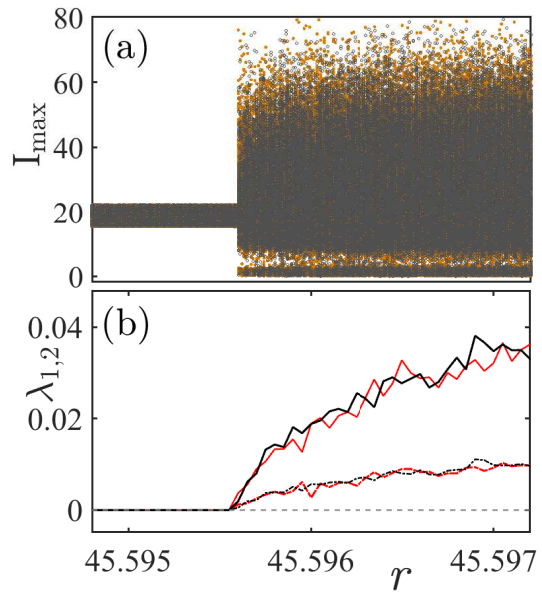}\\	
	\caption{Quasiperiodic intermittency to hyperchaos. Bifurcation diagram of $I_{max}$ in (a)  and a plot of two largest Lyapunov exponents $\lambda_{1,2}$ in (b) against $r$ in a range $r \in (45.594,45.598)$. Two Lyapunov exponents $\lambda_{1,2}$ are zero until $r=45.5956$ when two largest Lyapunov exponents suddenly become positive that indicates origin of hyperchaos.  A large expansion of $I_{max}$ occurs at this transition point $r=45.5956$ (a), which is a signature of LIE as manifested in the time evolution of $I$ in Fig.~\ref{fig5}(a).  Yellow dots and gray circles in (a), and red and black lines in (b) indicate forward and backward integrations of the model that show no shift in the transition point and confirm a hysteresis-free transition to hyperchaos at the critical  point. Other parameters are $\alpha$ =7.0, $\sigma$ = 6.0.}
	
	\label{fig4}
\end{figure}
\par Figures~\ref{fig3}(a) and \ref{fig3}(b) show the time evolution of laser intensity (I) and the phase portrait of $E_x^2$ against $E_y^2$, respectively,  near $r = 39.2142$ before the transition to hyperchaos at a critical point $r = 39.2141$. The dynamics is in the chaotic regime. The time evolution and the phase portraits of the dynamics after the transition to hyperchaos are demonstrated in Figs.~\ref{fig3}(d) and \ref{fig3}(e), respectively, show occasional large-intensity pulses or events away from the bounded region of attractor. A few intermittent large pulses are really very large and crosses a threshold height $h_s=\langle I_{max} \rangle$ +$6\sigma_I$ (horizontal dashed red line) where $\langle. \rangle$ denotes a mean and $\sigma_I$ defines the standard deviation. We call these extremely large-intensity pulses as LIE that have been triggered exactly when the transition to hyperchaos occurs.
\par  The probability distribution function (PDF) of all the peaks ($I_{max}=I_n$), small to large, is plotted in Fig.~\ref{fig3}(c) for the chaotic motion before the transition (for $r=39.2142$)  that shows an upper bound much below the $h_s$ line, which becomes extended with a tail beyond the threshold $h_s$ (vertical dashed line) in Fig.~\ref{fig3}(f) for $r= 39.2141$ when hyperchaos appears with a manifestation of LIE.  PDF of intensity pulses decreases with their height indicating rare occurrence of LIE beyond the $h_s$ line (vertical dashed line). For plotting all the PDF, the $t$-span length is chosen as $5.0\times 10^{9}$. In addition, we have plotted the inter-event interval (IEI) distribution for $r= 39.2141$ in Fig.~\ref{fig3}(g). The IEI distribution in the log-linear scale exhibits Poisson-like distribution that confirms the uncorrelated nature of the rare large-intensity pulses. The dashed red line Fig.~\ref{fig3}(g) denotes the exponential decay function $P(I)=ae^{-bI}$ and the fitting parameter values are $a = 0.1103$ and $b = 3.729\times 10^{-5}$.


\section{Hyperchaos: Quasiperiodic intermittency}
\par
\begin{figure}
	\includegraphics[width=0.58\columnwidth]{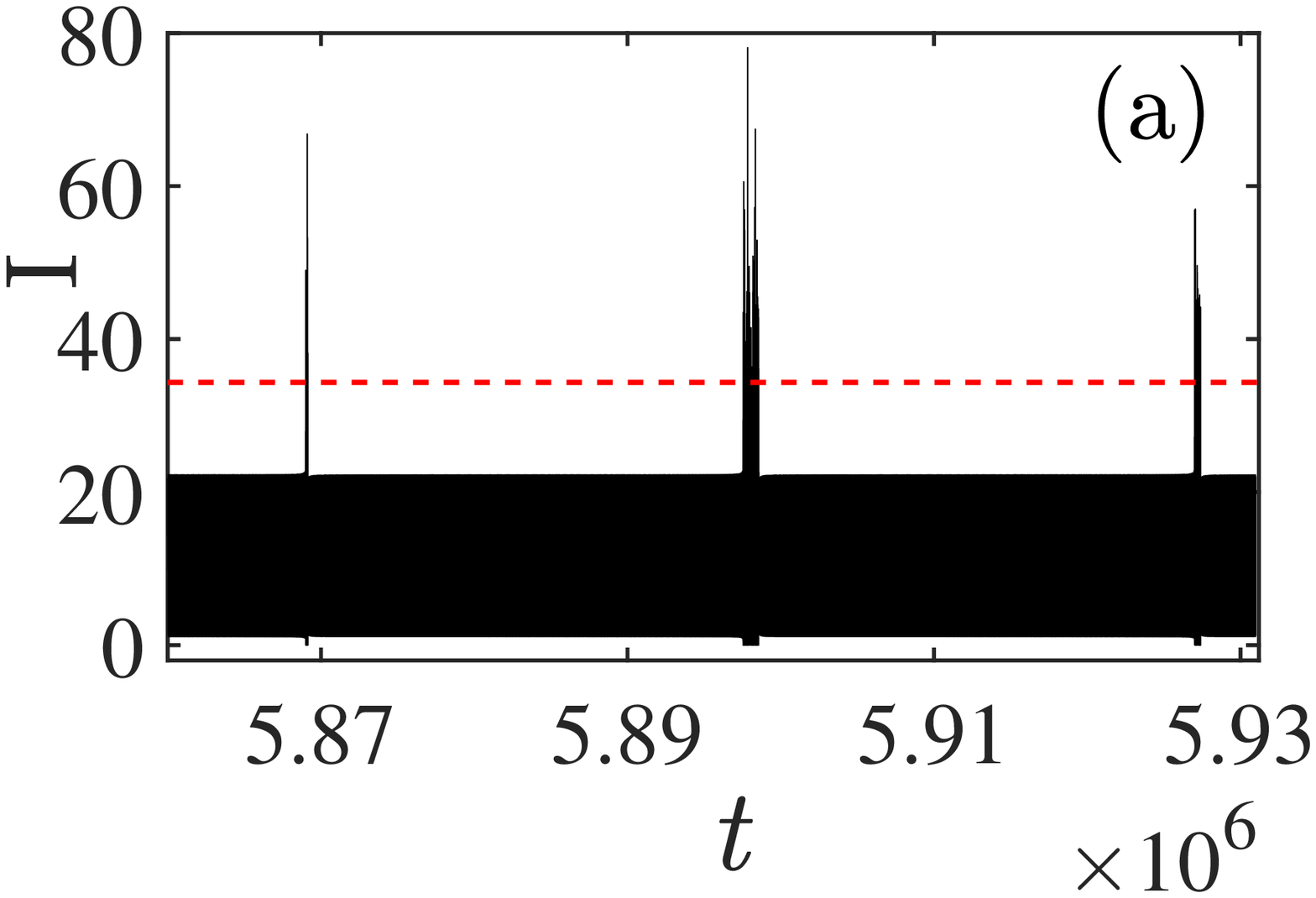}~\includegraphics[width=0.4\columnwidth]{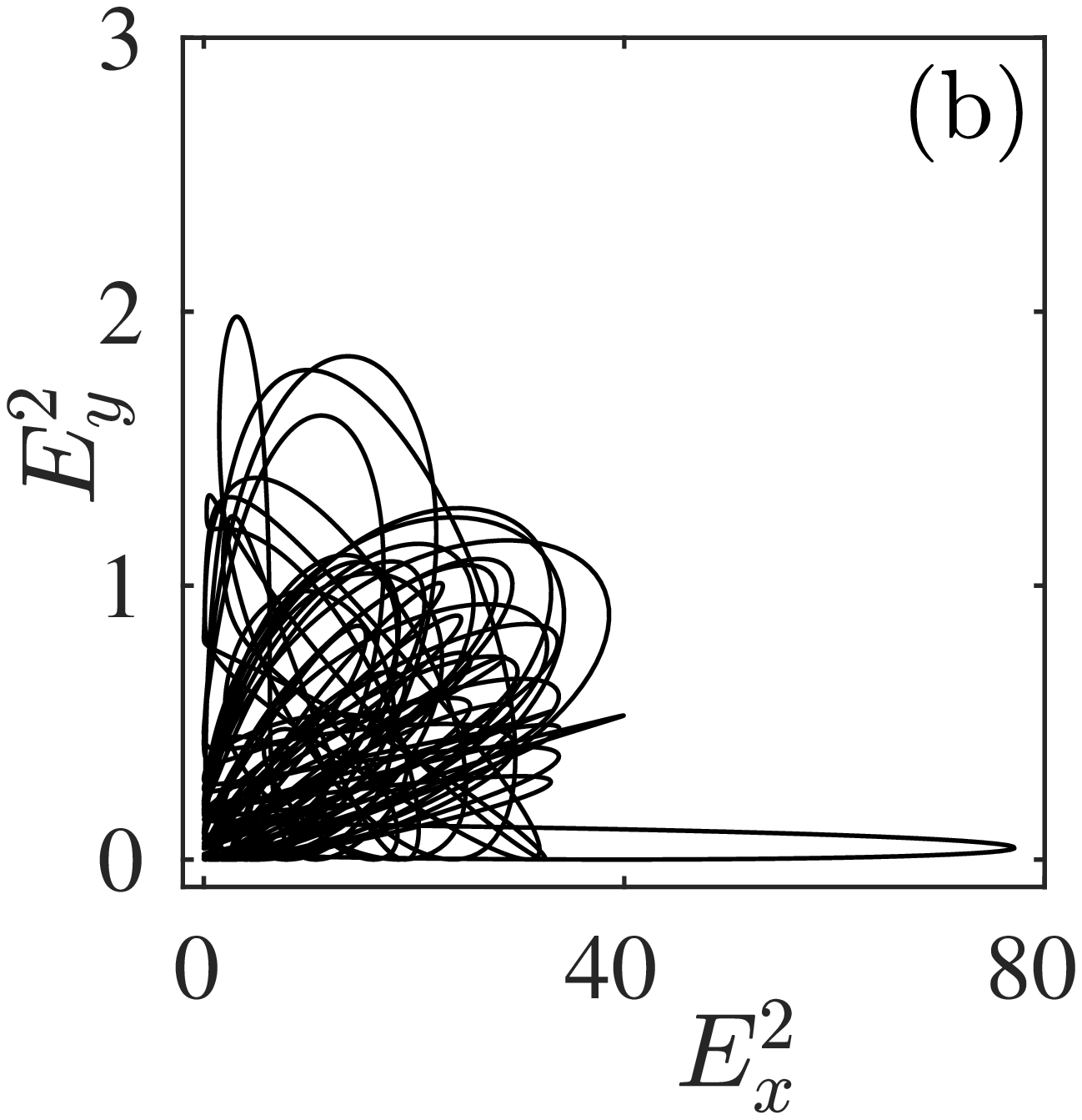}
	\includegraphics[width=0.45\columnwidth]{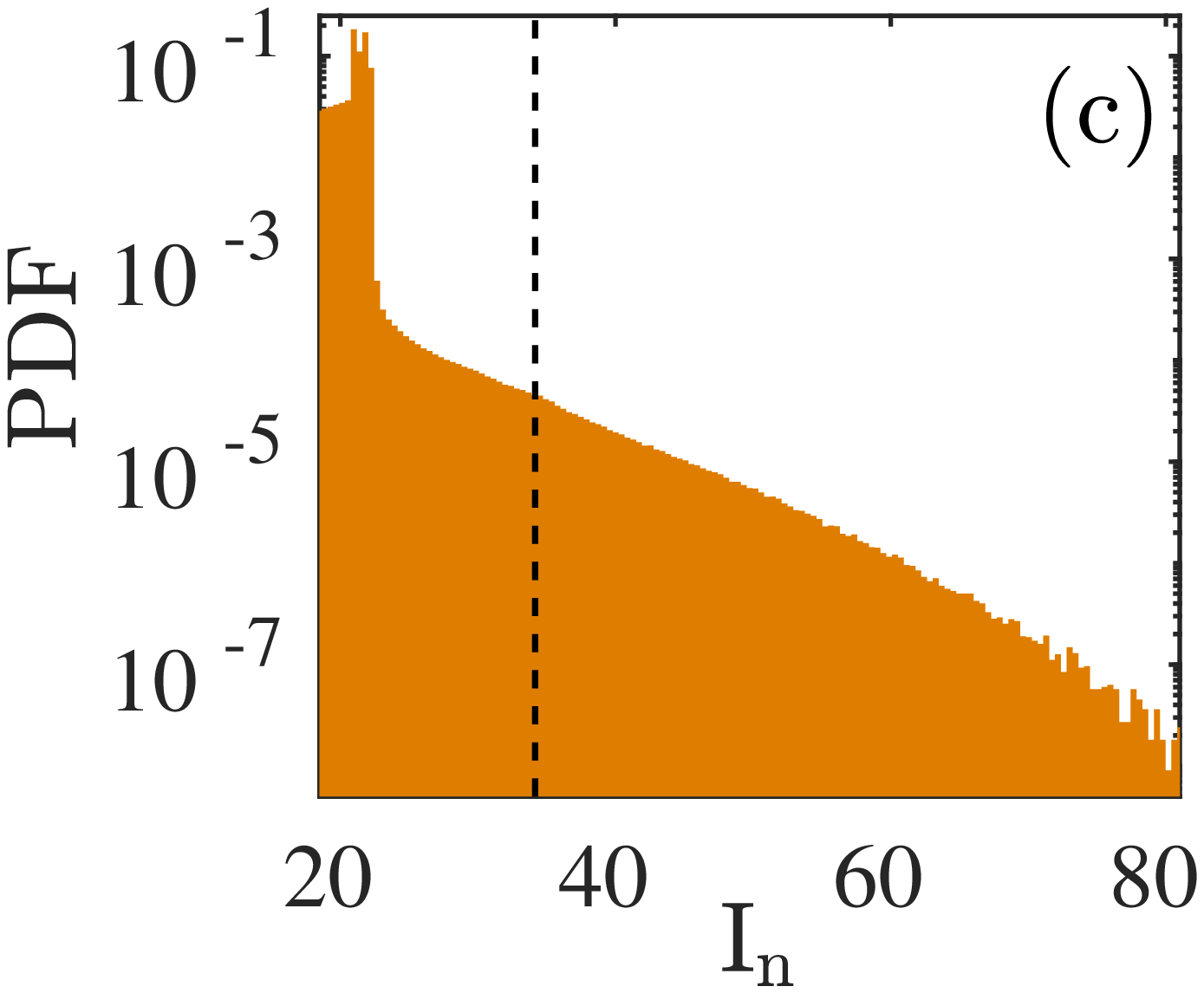}~\includegraphics[width=0.5\columnwidth]{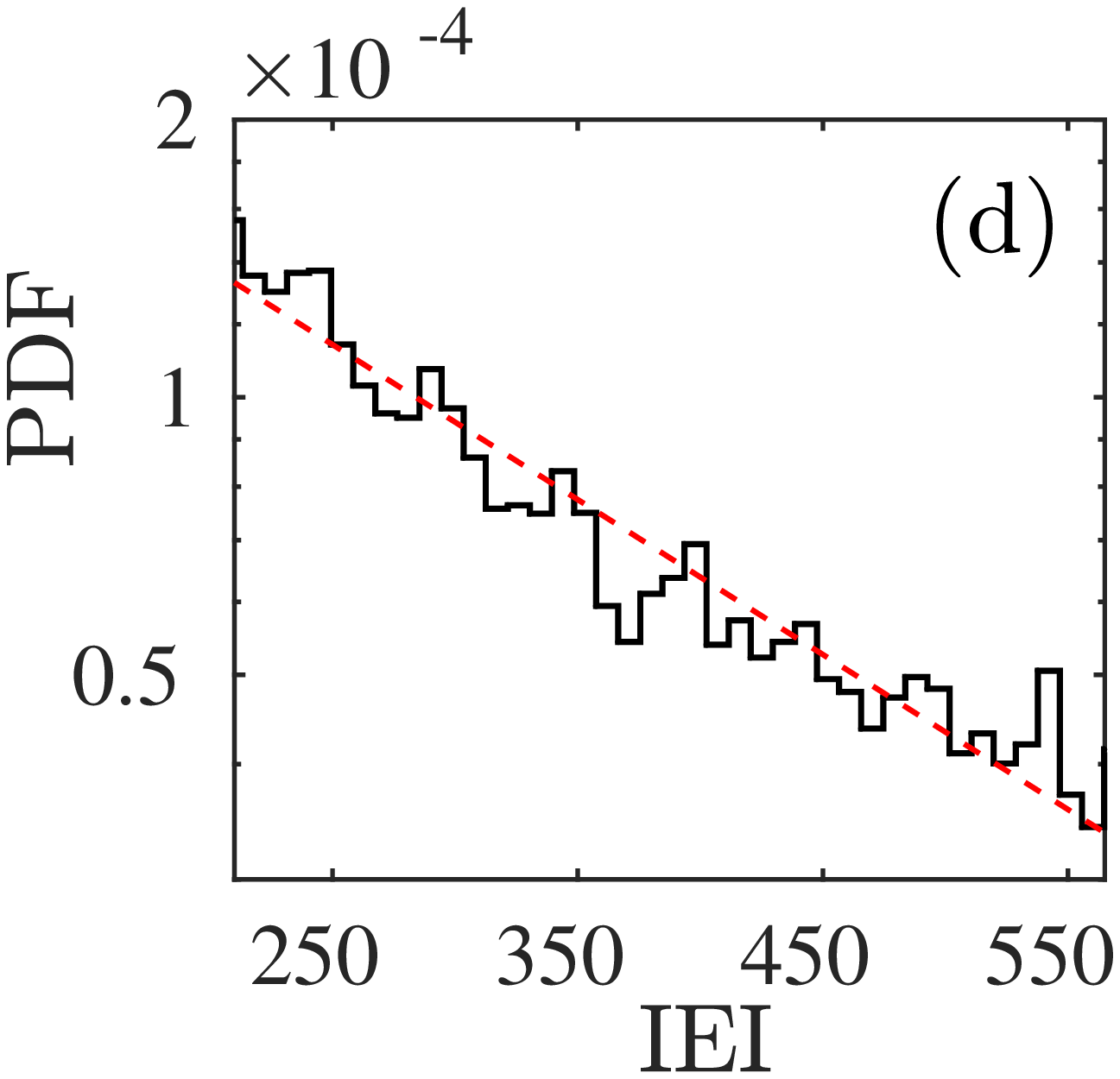}
	\caption{Hyperchaos via quasiperiodic intermittency. (a) Temporal evolution of $I$, (b) phase portrait of $E_x^2$ against $E_y^2$, and (c) PDF of local maxima $I_n=I_{max}$ for critical parameter $r= 45.5956$. A heavy-tail distribution of $I_n$ is seen confirming rare occurrence of LIE. The horizontal dashed line in (a) and vertical dashed line in (c) denote threshold height $h_s = \langle I_{n} \rangle$ + $6\sigma_I$.  (d) Probability distribution of IEI fitted with an exponential red dashed line.}
	\label{fig5}
\end{figure}

 Quasiperiodic intermittency as a source of instability is reported recently in the Zeeman laser \cite{redondo1997intermittent,redondo1996off, kingston2021instabilities} in relation to the origin of LIE. Intermittent large size chaotic bursts intercept the quasiperiodic motion in response to changes in the pumping parameter. It is similar, in some sense, yet different from the PM intermittency, since it originates from quasiperiodic motion instead of a periodic motion. We provide evidence  here that quasiperiodic intermittency also leads to hyperchaos when LIE starts appearing in the time evolution of laser intensity $I$. The bifurcation diagrams of $I_{max}$ and $\lambda_{1,2}$ are plotted along the solid black line in Fig.~\ref{fig1}(a), which are demonstrated in Fig.~\ref{fig4}. The bifurcation of $I_{max}$ in Fig.~\ref{fig4}(a) reveals quasiperiodic motion for $r < 45.5956$ when two largest Lyapunov exponents ($\lambda_{1,2})$ are zero as shown in Fig. \ref{fig4}(b). For increasing $r$ values, the system reveals a sudden large expansion in $I_{max}$  at a critical $r \approx 45.5956$ when $\lambda_{1,2}$ becomes both positive simultaneously from zero. Figures~\ref{fig4}(a)-(b) are drawn for both forward and backward integrations against $r$ as marked by two different colors showing no shift in the transition point. It clearly indicates that the transition from quasiperiodic motion to hyperchaos is discontinuous at the critical parameter $r \geq 45.5956$ and it is hysteresis-free. 
The time evolution of laser intensity (I) and the phase portrait of $E_x^2$ against $E_y^2$ at the origin of LIE or the hyperchaotic state ($r=45.5956$) are plotted in Figs.~\ref{fig5}(a) and \ref{fig5}(b). The time evolution shows longer laminar phases of low intensity quasiperiodic motion (not discussed in detail) occasionally intercepted by large-intensity pulses (turbulent phase). The phase portrait shows occasional journey of the trajectory far away from the dense region. The large-intensity spikes are much larger that the $h_s$ line (horizontal red dashed line). 
The probability distribution function (PDF) plot of all the intensity pulses in a long time series of $I$  during hyperchaos shows a heavy-tail thereby confirming rare appearance of LIE in  Fig.~\ref{fig5}(c).  Furthermore, the IEI distribution fitted by an exponential function $P(I)=ae^{-bI}$ is depicted in Fig.~\ref{fig5}(d), where its parameters  are $a = 3.0\times 10^{-4}$ and $b = 0.0039$.  Quasiperiodic intermittency, thus, leads to the origin of hyperchaos at a critical parameter concurrently with the appearance of rare and recurrent LIE. 


\section{Hyperchaos: PM intermittency}
\begin{figure}
	\includegraphics[width=0.85\columnwidth]{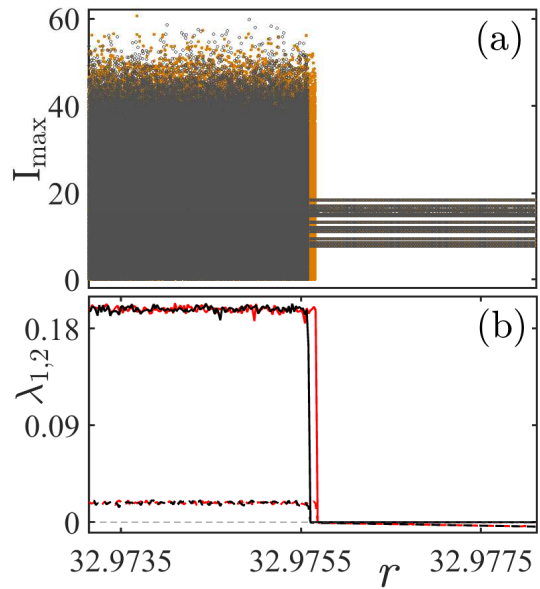}\\	
	\caption{Hyperchaos and LIE for PM intermittency. (a) Bifurcation of  $I_{max}$ and (b) a plot of two largest Lyapunov exponents $\lambda_{1,2}$ against $r \in (32.973, 32.978)$. Yellow dotes and black circles in (a), and red and black lines in (b) indicate forward and backward integrations that reveal a shift in the transition point confirming the presence of hysteresis at the critical point. At critical  $r \approx$  32.9755 (a period-14 orbit transits to LIEs), the two largest Lyapunov exponent become positive that indicates origin of hyperchaos.} 
	\label{fig6}
\end{figure}
Finally, we elaborate how PM intermittency \cite{pomeau1980intermittent} leads to hyperchaos with the characteristic feature  of LIE in the temporal dynamics of the laser. A different set of parameters $\alpha$ = 4.1, $\sigma$ = 6.0 are chosen and  the pumping rate $r$ is varied in the range $r \in (32.973, 32.978)$ following a very slim dotted line marked in Fig.~\ref{fig1}(b). 
The bifurcation diagram  $I_{max}$ in Fig.~\ref{fig6}(a) plotted against $r$ manifests a transition from a periodic state (period-14) to hyperchaos in a discontinuous fashion at a critical $r \approx 32.9755$.  The periodic nature of the dynamics before the transition is confirmed by $\lambda_2$ (dashed line) which continues to be negative while $\lambda_1=0$ until a critical point of transition appears when $\lambda_{1,2}$ both become positive confirming the origin of hyperchaos directly from a periodic state. However, the transition point shifts with forward and backward integration of the system as depicted by two different bifurcation diagrams in colors (gray and yellow) and also from $\lambda_{1,2}$ plots (black and red lines) against $r$ in Fig.~\ref{fig6}(a) and \ref{fig6}(b), respectively. Thus the transition from periodic to hyperchaos via PM intermittency shows a hysteresis. 
Nevertheless, the appearance of hyperchaos and recurrent LIE  exactly coincide at the critical parameter point. For a lower value of $r \le$ 32.9755, LIE continues to exist. The magnitude of two largest Lyapunov exponents (both $\lambda_1$, and $\lambda_2$) remains positive and almost constant with respect to the pumping parameter. 
\begin{figure}
\includegraphics[width=0.55\columnwidth]{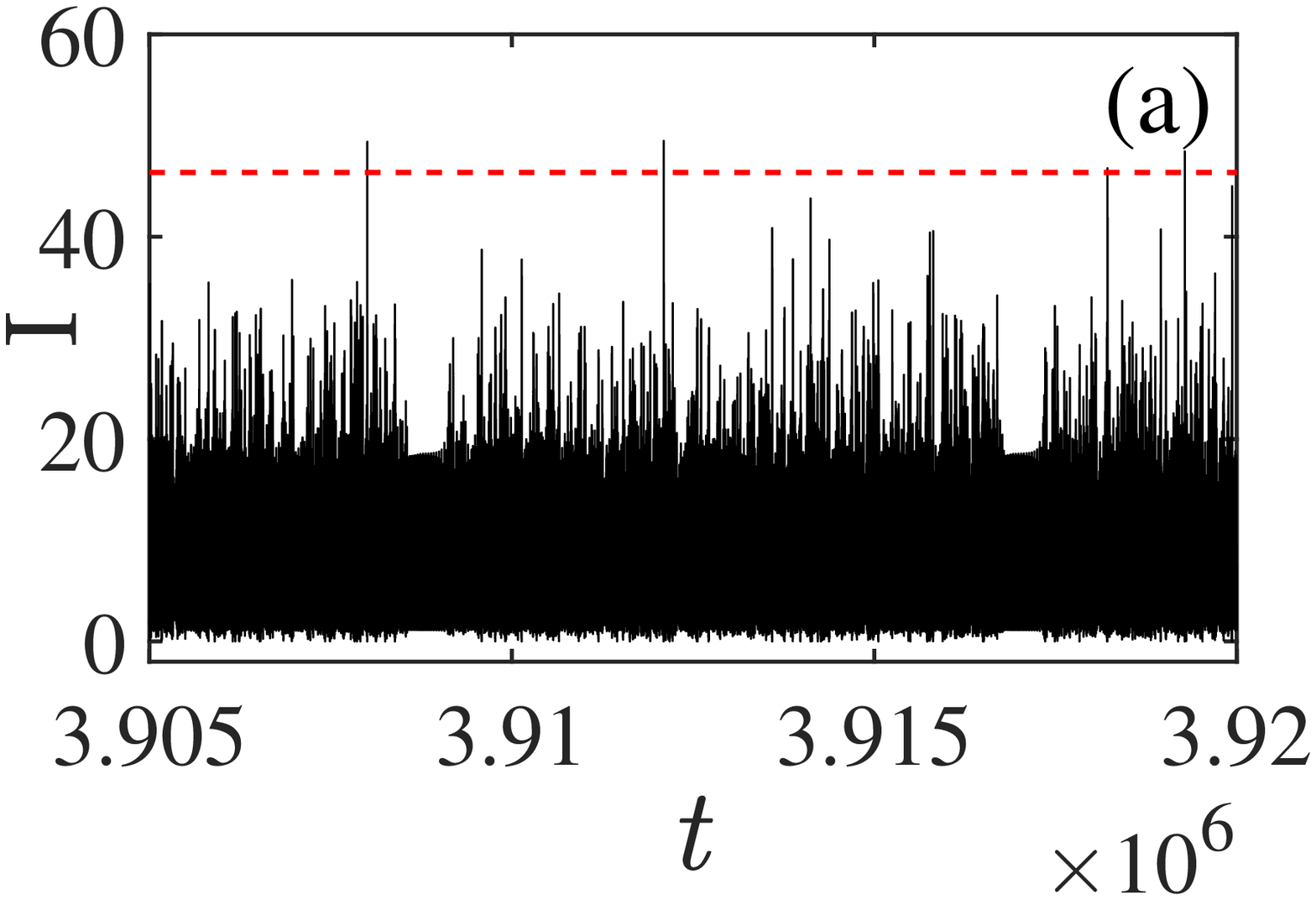}~\includegraphics[width=0.45\columnwidth]{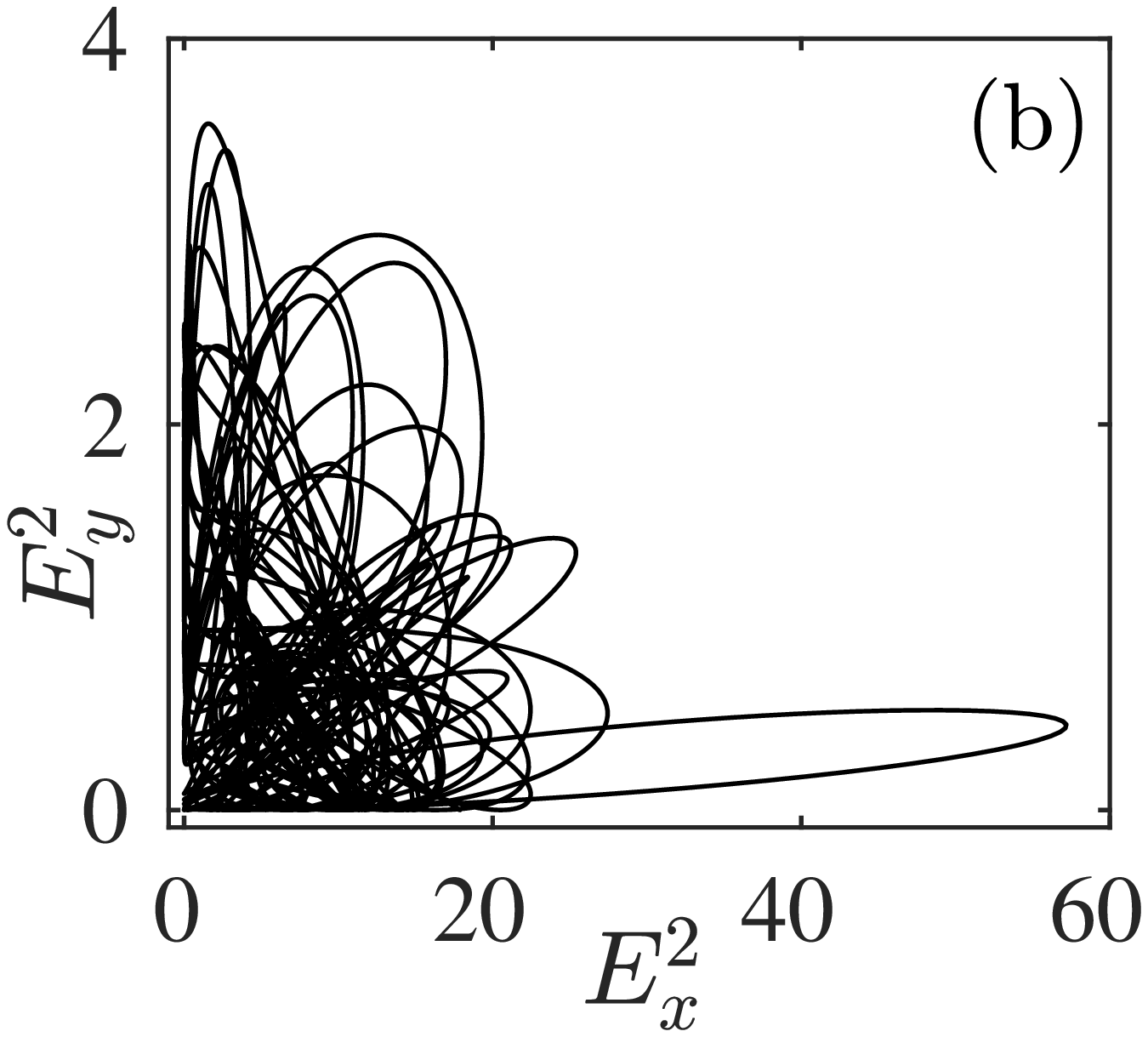}\\

\includegraphics[width=0.5\columnwidth]{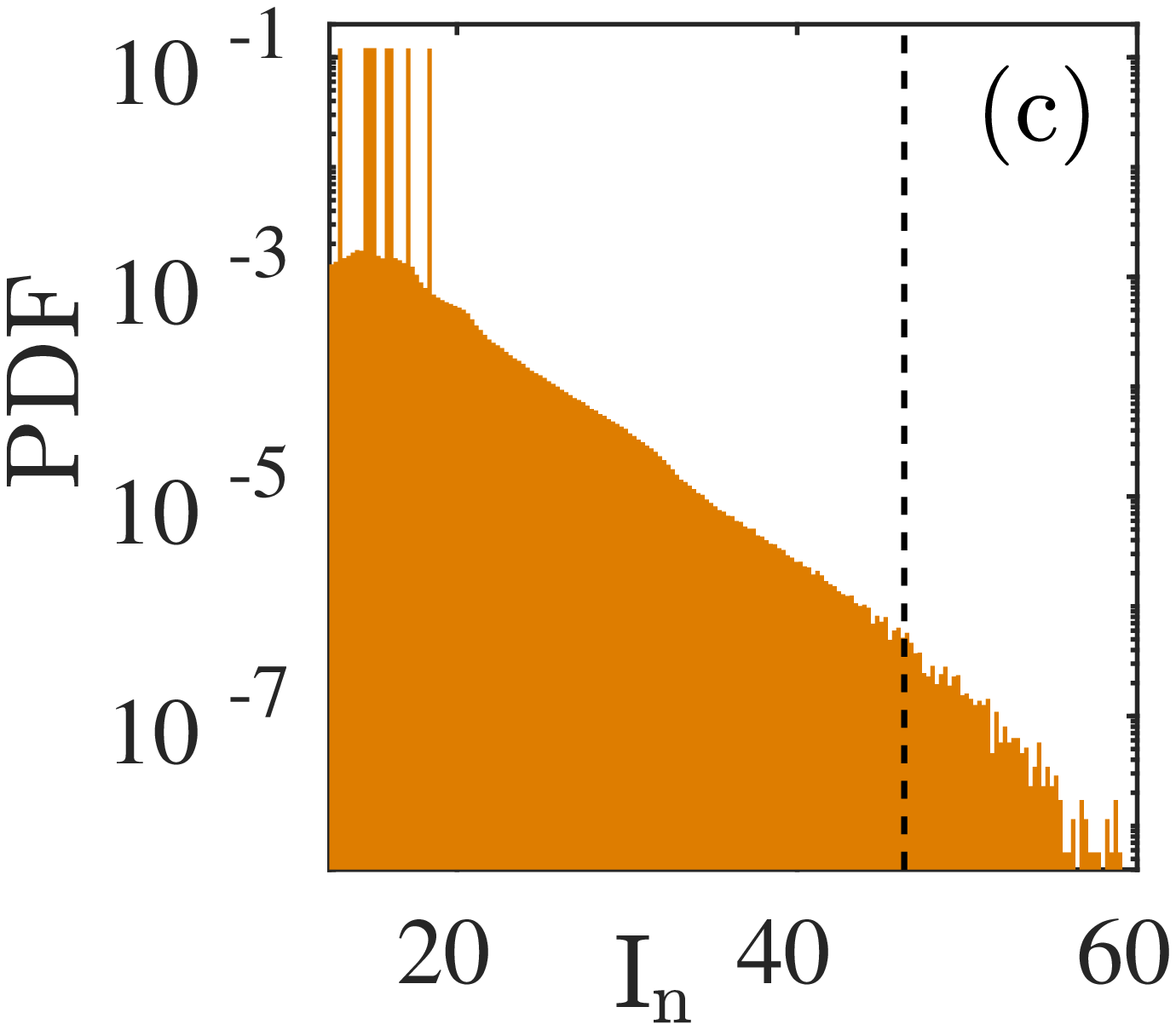}~\includegraphics[width=0.5\columnwidth]{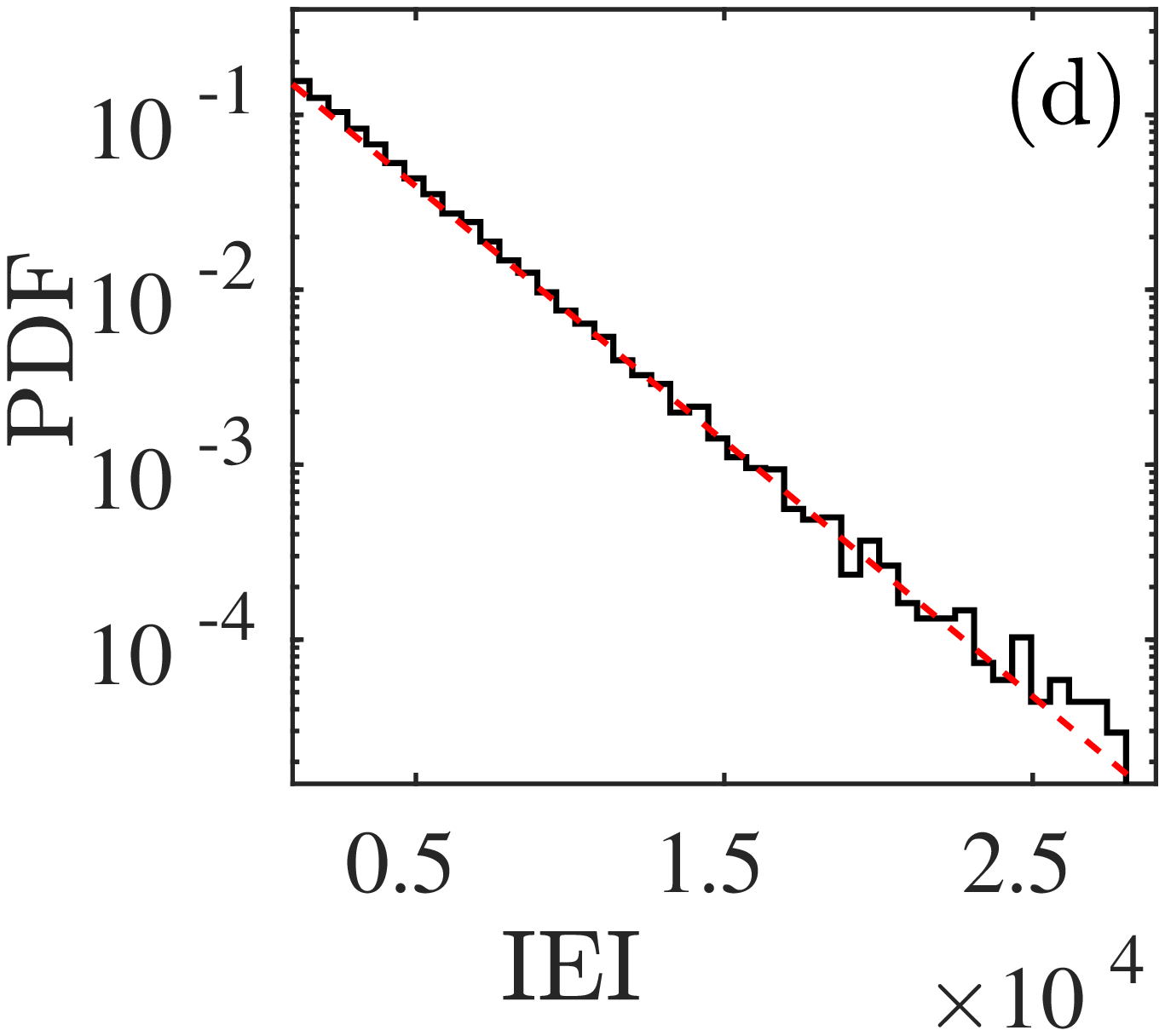}
	\caption{PM intermittency and origin of hyperchaos. Time evolution of $I$ in (a) and PDF of local maxima $I_n$ in (b) for $r \approx 32.9755$. Rare large-intensity pulses are seen to cross the horizontal dashed line (threshold height $h_s$) in the time evolution of $I$. (c) PDF of $I_n$ confirms rare occurrence of LIE beyond the threshold $h_s$ line (vertical dashed line). (d) PDF of IEI fitted with an exponential line (red dashed line).}
	\label{fig7}
\end{figure} 
We check the temporal dynamics and phase portrait of hyperchaos, once again, which confirms  the appearance of LIE in Fig.~\ref{fig7}(a) and \ref{fig7}(b), which are recurrent and indicate their occasional far away journey from the dense attractor. The rare occurrence of LIE is manifested in the PDF with a tail beyond the threshold $h_s$ line (vertical dashed line) in Fig.~\ref{fig7}(c). The  distribution of IEI that appears infrequently from the laminar phase of periodic motion portrayed in Fig.~\ref{fig7}(d) manifests a Poisson-like distribution which is also fitted by exponential function with $a = 0.2087$ and $b = 3.356\times 10^{-4}$.  

\section{Effect of Noise}
We study the effect of noise on the origin of hyperchaos with the addition of noise in the dynamics of $E_x$ in Eq.(1),  
\begin{eqnarray}
\dot{E}_x &=&\sigma(P_x - E_x)+\sqrt{D}\xi, 
\label{Zee:eqn_noise}
\end{eqnarray}	
where $\xi$ denotes Gaussian noise with zero mean and unit variance and $\mathrm{D}$ is its relative intensity. We  solve the noise-induced Zeeman laser model  using the modified Runge-Kutta method based on Ito's algorithm Ref.\cite{saito1996stability,newton1991asymptotically}.
\begin{figure}   
	\includegraphics[width=0.525\columnwidth]{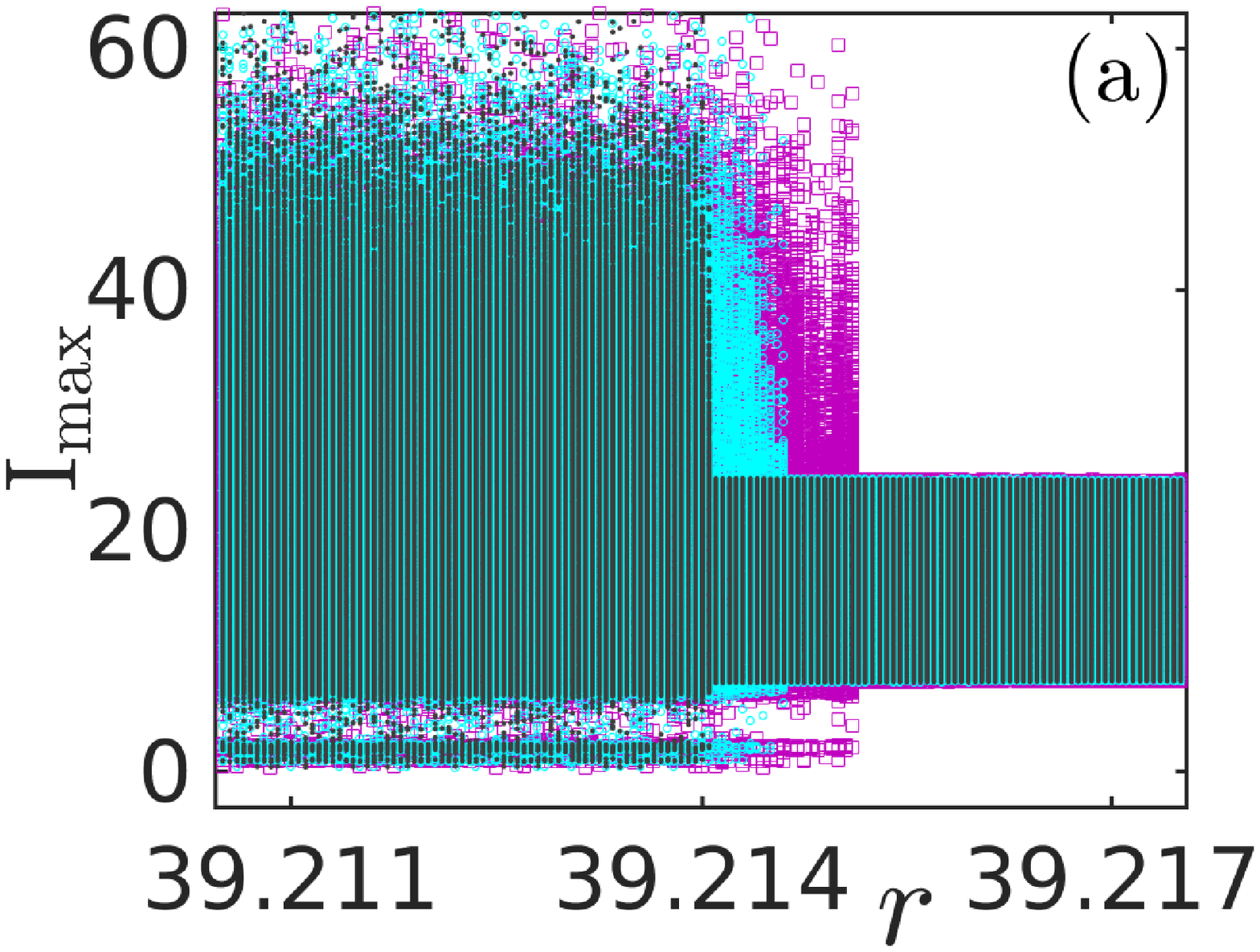}~\includegraphics[width=0.425\columnwidth]{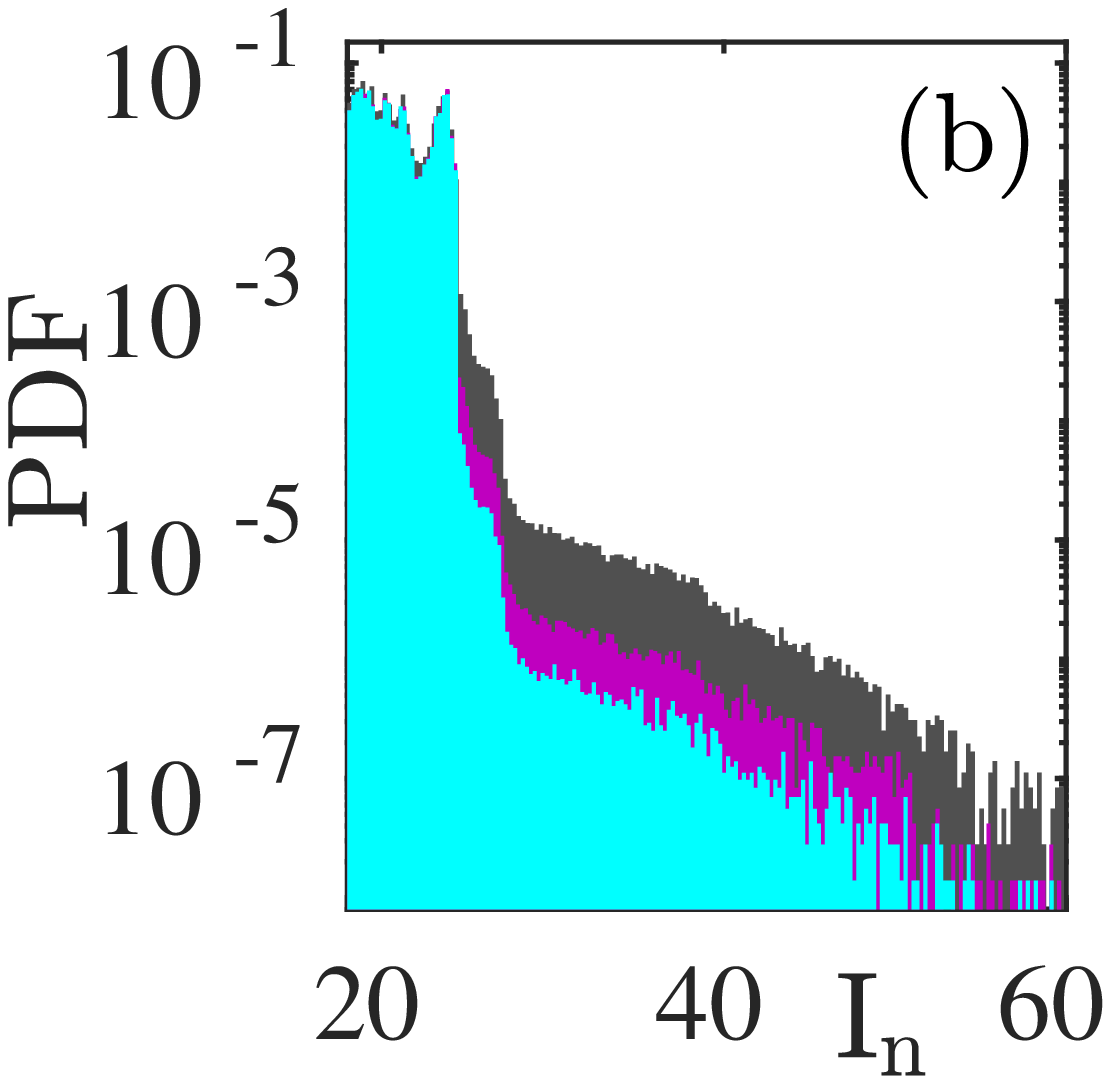}	
	\caption{Noise effect on transition to hyperchaos: quasiperiodic breakdown to chaos followed by crisis and origin of LIE. (a) Bifurcation diagrams of the local maxima of laser intensity ($I_{max}$)  against the pumping parameter $r \in$ (39.2105, 39.2175) for $D = 0.0$ (gray), $D = 1.0\times10^{-4}$ (cyan), and $D = 2.0\times10^{-4}$ (magenta). (b) PDF of intensity events ($I_{max}$) for different noise strength and pumping parameter values, for $D = 0.0$, $r$ = 39.2141 (gray), $D = 1.0\times10^{-4}$, $r$ = 39.2146 (cyan), and $D = 2.0\times10^{-4}$, $r$ = 39.2152 (magenta), respectively.}
	\label{fig8}
\end{figure}  
\par First, we elucidate the noise effect during the quasiperiodic breakdown to chaos followed by crisis with  bifurcation diagrams against the system parameter $r$ for two different noise intensities (cyan and magenta colors) as presented in Fig.~\ref{fig8}(a) and compare with the noise-free condition (gray color). Three bifurcation diagrams are placed on top of each other using different colors for different strength of noise, $D = 0.0$ (gray dots), $D = 1.0\times10^{-4}$ (cyan circles), and $D = 2.0\times10^{-4}$ (magenta squares), respectively. The critical point of transition to hyperchaos clearly shifts with noise strength, however, the characteristic  feature of the discontinuous transition with a large expansion of the attractor still persists. We do not present the temporal dynamics, however, the PDF of ($I_n=I_{max}$) for different noise strength is depicted in Fig.~\ref{fig8}(b). Different $r$ values are chosen for the plot since the transition point shifts with noise strength. 
The statistics of non-Gaussian distribution  with a tail remains unchanged (cyan and magenta) when compared with the  distribution ($D$ = 0.0, gray color) in absence of noise. A little shift in the values of PDF is only reflected for increasing noise. This confirms rare occurrences of LIE that continues to appear even in the presence of noise.
\begin{figure}   
	\includegraphics[width=0.55\columnwidth]{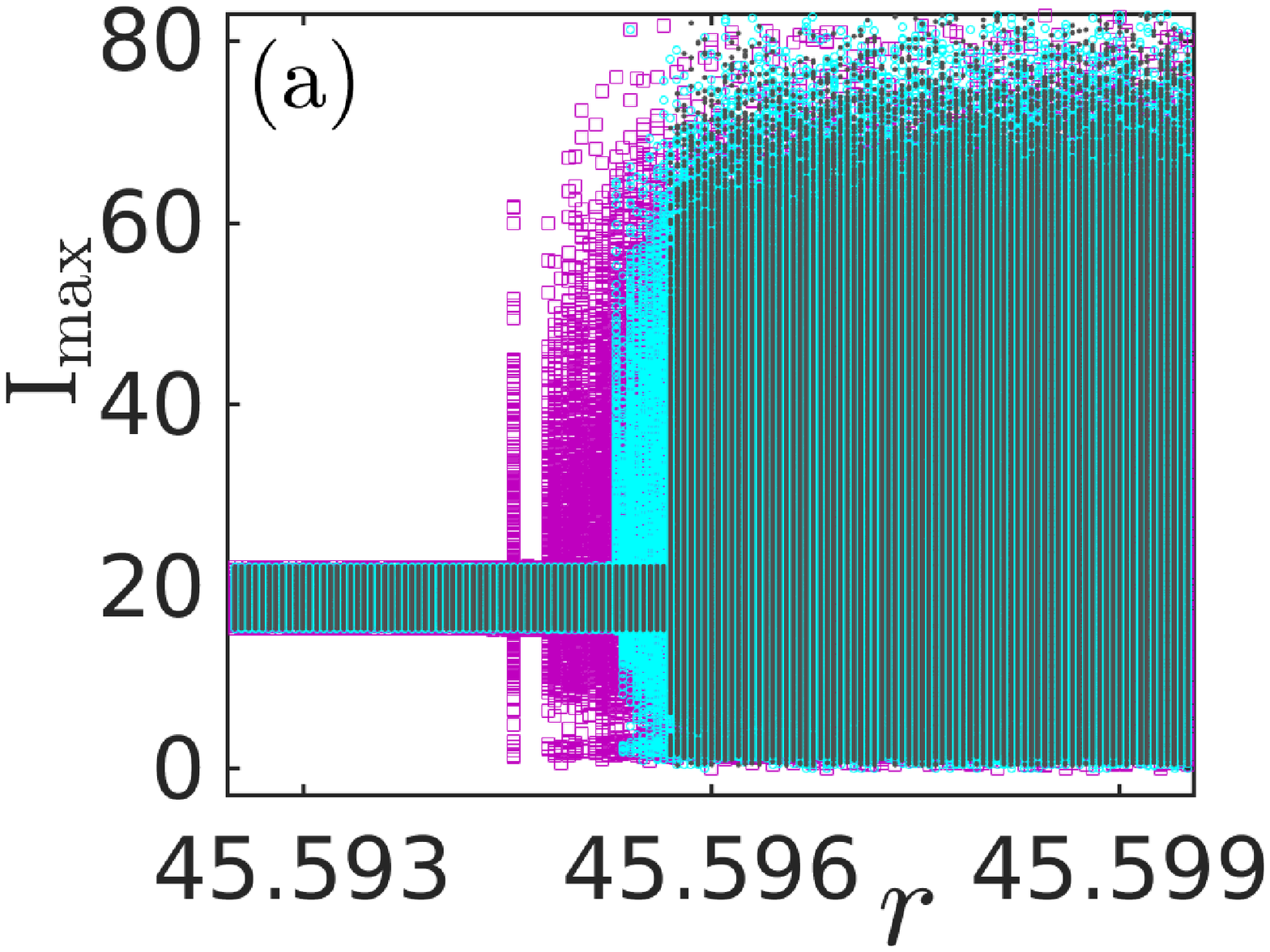}~\includegraphics[width=0.45\columnwidth]{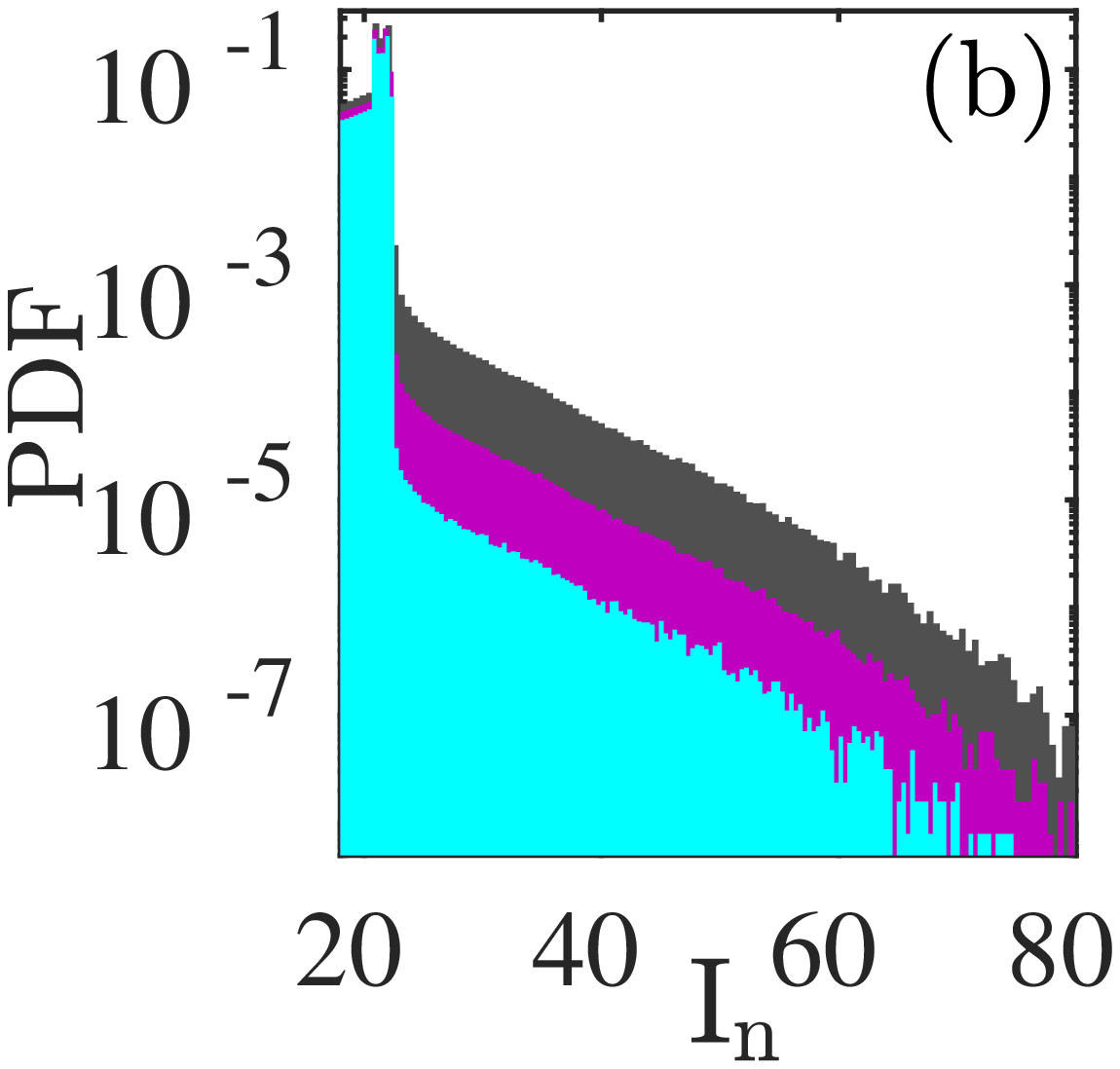}	
	\caption{Noise effect during a transition to quasiperiodic intermittency. (a) Bifurcation diagrams of $I_{max}$ plotted against $r$ $\in  (45.5925, 45.5995)$ for  $D = 0.0$ (gray), $D = 5.0\times10^{-4}$ (cyan), and $D = 1.0\times10^{-3}$ (magenta). (b) PDFs for different noise strength and pumping parameter values, for $D = 0.0$, $r$ = 45.5956 (gray), $D = 5.0\times10^{-4}$, $r$ = 45.5953 (cyan), and $D = 1.0\times10^{-3}$, $r$ = 45.5949 (magenta).}
	\label{fig9}
\end{figure} 
\par We observe a similar noise effect  during the quasiperiodic intermittency. The bifurcation diagram for $I_{max}$ against the pumping parameter $r \in$ (45.5925, 45.5995) is portrayed in Fig.~\ref{fig9}(a). A clear view of the shift in the transition point is obtained against  noise strength. 
PDF of ($I_{max}$) for different noise strength are shown in Fig.~\ref{fig9}(b), where it shifts to lower values (cyan and magenta colors) from the noise-free condition (gray color). The heavy-tail characteristics of the distribution of $I_{max}$ remains.
\begin{figure}   
	\includegraphics[width=0.55\columnwidth]{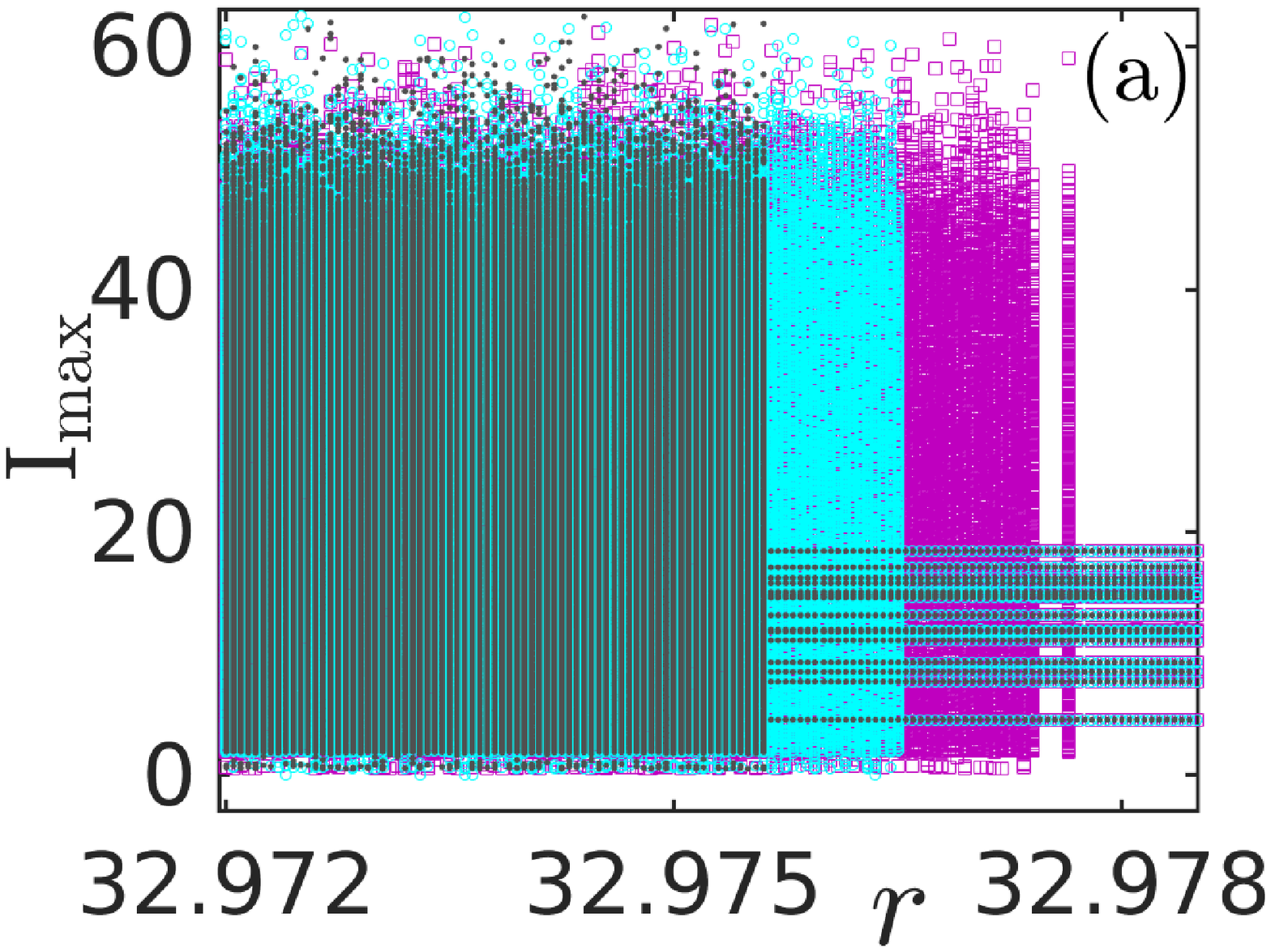}~
	\includegraphics[width=0.45\columnwidth]{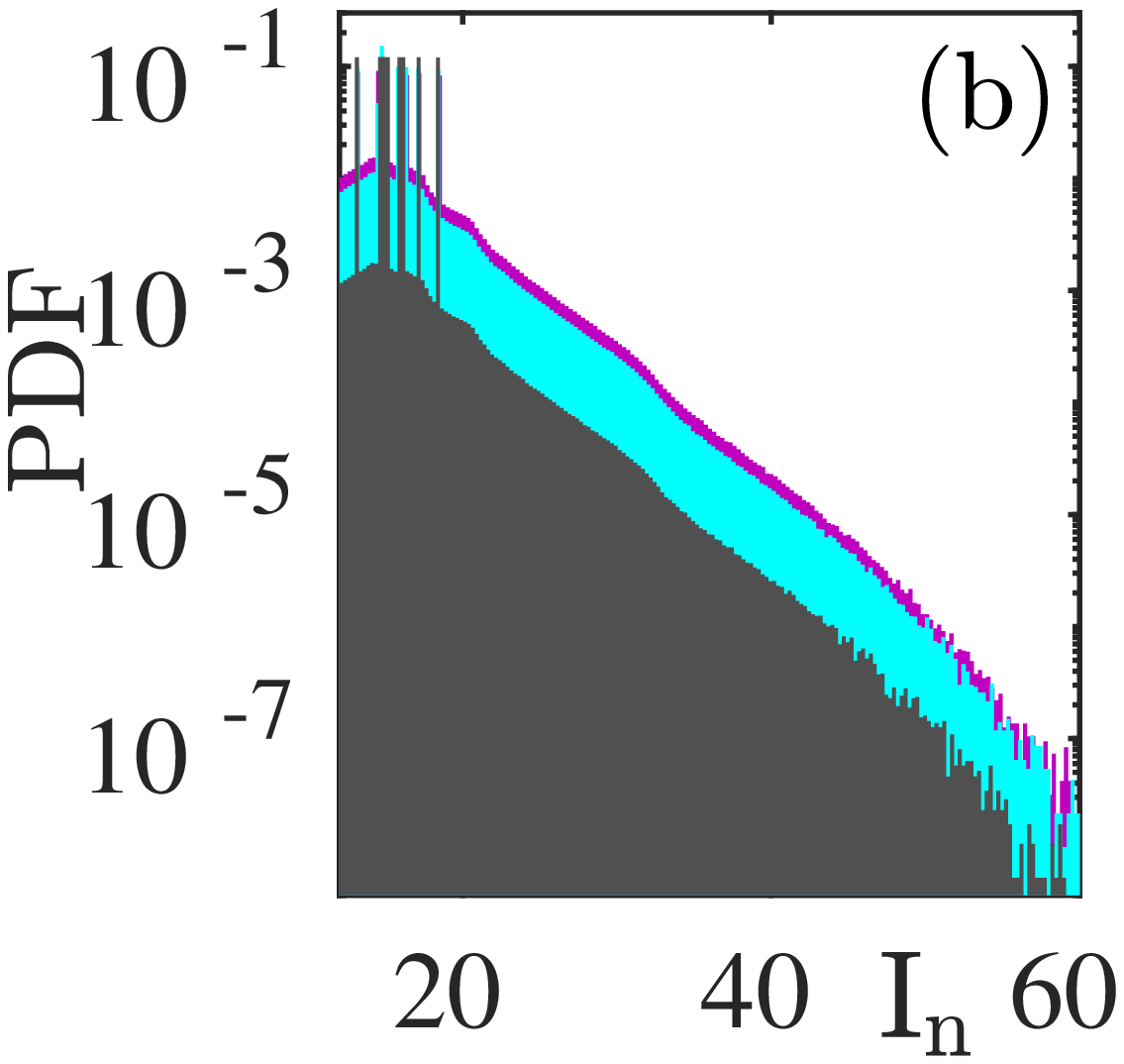} \\
	\caption{Noise effect during PM intermittency transition. $I_{max}$ against $r$ $\in (32.972, 32.979)$ for $D = 0.0$ (gray), $D = 1.0\times10^{-4}$ (cyan), and $D = 2.0\times10^{-4}$ (magenta). (b) PDF for different noise strength and pumping parameter values, for $D = 0.0$, $r$ = 32.9755 (gray), $D = 1.0\times10^{-4}$, $r$ = 32.9765 (cyan), and $D = 2.0\times10^{-4}$, $r$ = 32.9774 (magenta).}
	\label{fig10}
\end{figure}  
\par During the PM intermittency route to hyperchaos, the bifurcation diagrams are shown in Fig.~\ref{fig10}(a) as plotted for $D$ = 0.0 (gray),   $D = 1.0\times10^{-4}$ (cyan),  and  $D = 2.0\times10^{-4}$ (magenta). Once again, it confirms that the critical parameter point shifts in response to the noise strength. PDF of ($I_{max}$) shows a shift to higher probabilities under the influence of noise as illustrated in Fig.~\ref{fig10}(b), however, the basic character of the distribution remains unchanged and confirms rare occurrences of LIE. 
\\
\section{conclusion}
We have explored the Zeeman laser model and found three sources of instabilities, breakdown of quasiperiodicity to chaos followed by interior crisis, quasiperiodic intermittency, and PM intermittency. The prescribed instabilities originate extreme events at discrete critical parameters in the laser model, for all three cases, in response to a change in the pumping parameter.  We have now recognized the dynamic nature of extreme events denoted here as LIEs as hyperchaotic. Hyperchaos has been recognized by estimating the Lyapunov exponents of the system when two of them become positive.  The temporal dynamics of hyperchaos in all three cases show LIE with characteristic features of extreme events, in the sense, that intermittent large events are larger than a threshold height and they appear on rare occasions, which has been confirmed by the non-Gaussian distribution with a tail (light or heavy) of all the intensity pulses. On the other hand, the IEI distribution manifest a Poisson-like process.  A common scenario of a discontinuous large expansion of the attractor of the laser model is noted during the appearance of hyperchaos for all the three cases. The transition to hyperchaos and the appearance of LIE are  concurrent and discontinuous for all the three cases, and hysteresis-free except in the case of PM intermittency when the transition  shows a shift in the critical point with forward and backward integrations. 
\par We have noticed a shift in the transition points to hyperchaos against the parameter shifts in the presence of weak noise for all the three cases, however, the fundamental feature of a discontinuous large expansion of the attractor  with the origin of LIE remains unchanged. Extreme events-like LIE added more complexity in the dynamics of the laser model than typical chaos by occasionally deviating from it and producing recurrent large-intensity  events and are now identified as hyperchaotic. Besides the laser model, this phenomenon of the origin of hyperchaos and LIE is common as usually found \cite{kingston2022transition} in other paradigmatic models. This observation raises a question about the true dynamical character of extreme events as observed in low dimensional systems. Extreme events, in general, in dynamical systems, low or high dimensional, are more complex than simple chaos. We need a new metric to define the complexity of extreme events in any model systems.
\begin{acknowledgments}
	T.K., and S.L.K. have been supported by the National Science Centre, Poland, OPUS Programs (Projects No. 2018/29/B/ST8/00457, and 2021/43/B/ST8/00641). S.L.K acknowledges PLGrid Infrastructure (Poland) for the computation facility. M.B. has been supported by the National Science Centre, Poland under project no. 2017/27/B/ST8/01619. S.K.D. acknowledges financial support from the Division of Dynamics, Lodz University of Technology, Lodz, Poland.   
\end{acknowledgments}     
\section*{Data Availability}
The data that support the findings of this study are available from the corresponding author upon request.
\section*{References}
\bibliography{hyper_ref_fi}
\end{document}